\documentclass[11pt]{article}
\usepackage{latexsym}
\usepackage{amsmath}
\usepackage{amsfonts}
\usepackage{dsfont}
\usepackage{verbatim}
\usepackage{hyperref}
\usepackage{tikz}
\usetikzlibrary{matrix,arrows}
\usepackage{subcaption}
\usepackage{amsmath,amssymb}
\usepackage{graphicx}
\usepackage{color}
\usepackage{enumerate}
\usepackage{booktabs}
\usepackage{multirow}
\def\be{\begin{equation}}
\def\ee{\end{equation}}
\def\bea{\begin{eqnarray}}
\def\eea{\end{eqnarray}}

\begin{document}

\begin{flushright}

\end{flushright}

\vspace{40pt}

\begin{center}

{\Large \textbf{Shadows of Kerr-Vaidya-like black holes} } \\
\vspace{20pt}
Hai Siong Tan \\
\vspace{15pt}
Email: haisiong.tan@pennmedicine.upenn.edu\\
\vspace{15pt}
University of Pennsylvania, Perelman School of Medicine, Department of Radiation Oncology, Philadelphia, USA\\
\vspace{25pt}
%{\sc  Jan 2023}
\end{center}
\vspace{25pt}
\begin{center}
Abstract
\end{center}
In this work, we study the shadow boundary curves of rotating time-dependent black hole solutions which have
well-defined Kerr and Vaidya limits. These solutions are constructed by applying the Newman-Janis algorithm 
to a spherically symmetric seed metric conformal to the Vaidya solution with a mass function that is linear in Eddington-Finkelstein coordinates. Equipped with a conformal Killing vector field, this class of solution exhibits separability of null geodesics, thus allowing one to develop an analytic formula for the boundary curve of its shadow. We find a simple power law describing the dependence of the mean radius and asymmetry factor of the shadow on the accretion rate.  Applicability of our model to recent Event Horizon Telescope observations of M87${}^*$ and Sgr A${}^*$ is also discussed.

\newpage

\tableofcontents 

\section{Introduction}

Recent Event Horizon Telescope (EHT) observations of horizon-scale shadow images of M87${}^*$ \cite{EHTM871} and Sgr A${}^*$ \cite{EHTSgr1} have
furnished not only a direct visual evidence of black holes, but have also led to many new constraints on 
various potential deviations from General Relativity. The boundary curve of the black hole shadow emerges
from light rays that spiral asymptotically from the photon region demarcating the boderline between light rays that will eventually be captured by the black hole and those that escape to infinity \cite{Falcke}.\footnote{This curve is termed as the `critical curve' by Gralla et al. in \cite{Gralla} and `apparent boundary' by Bardeen in \cite{Bardeen}. See for example the review of \cite{Oleg} for an extensive discussion of basic ideas and history. } The geometry of this boundary curve depends on the background metric which could thus be probed by EHT observations \cite{EHTM875,EHTSgr6,Cheyu,Sunny1,Sunny2,Sunny3} . For example, the shadow geometry of 
Sgr${}^*$ has been used to exclude the central object being a Reissner-Nordstrom-type naked singularity or a traversable Misner-Thorne wormhole \cite{EHTSgr6}. 

Surrounding the black hole shadow is an emission ring of which structure is sensitive to a rich set of astrophysical phenomena, such as radiative transfer, that characterize the matter-energy accretion process. Typically, 
general relativistic magnetohydrodynamic (GRMHD) simulations are used to model the accretion flow processes 
\cite{EHTM875,EHTSgr6,EHTMag,EHTSgr5}. For the EHT experiments, they have revealed the emission ring properties to be consistent with a number of accretion models built upon the background of a Kerr black hole \cite{EHTM875,EHTSgr6}. The spacetime
metric in these simulations is assumed to be purely Kerr spacetime throughout, 
with the energy-momentum tensor capturing the magnetic field and average plasma properties \cite{EHTMag,McKinney}. 
In \cite{Johannsen,Chan,Narayan}, it was noted that the shadow size and shape is hardly influenced by the accretion details, and thus serves as a pristine signature of spacetime geometry. An implicit assumption is that the backreaction of the GRMHD energy-momentum tensor on the metric has a negligible influence on the shadow and could thus be ignored in deriving its geometry. 

In this paper, we study the shadow boundary curves of a class of rotating Kerr-Vaidya-like black hole solutions. 
Their metrics can be described as deformations of the Kerr solution parametrized by a small dimensionless parameter $\mu$. This parameter can be interpreted as the mass accretion rate constant in natural units. 
In the limit of vanishing spin, our spacetime reduces to a well-known model of spherically accreting black hole - the Vaidya spacetime
with a mass function $\mu v$, with $v$ being an ingoing Eddington-Finkelstein coordinate.
The latter solution was studied most recently in \cite{Solanki} where the authors derived and examined its shadow characteristics analytically. Most crucially, an analytic treatment of the shadow was possible by virtue of the existence of a Carter constant leading to separability of its null geodesic equations. This is related to a conformal Killing symmetry associated with the linear mass function, and hence its choice, for it enables the authors of \cite{Solanki} to derive explicit formulas for the radius of the photon sphere and the shadow angular diameter.  
One main motivation of our work here is to seek a rotating generalization of the analytic treatment in \cite{Solanki}. This would serve as a simple model of a backreacted Kerr-like geometry that is accreting mass, and for which an analytic derivation of its shadow geometry is possible. For readers familiar with exact solutions in GR, 
a natural candidate would be the Kerr-Vaidya solution \cite{Dahal} which can be obtained by replacing the constant Kerr mass with a variable mass function in the original Kerr line element expressed in Eddington-Finkelstein coordinates. Unfortunately, as we'll elaborate later, this solution does not offer any additional Carter constant that could lead to its null geodesic equations being separable.

We construct our solutions by applying the Newman-Janis algorithm \cite{Newman}
to a spherically symmetric seed metric conformal to the Vaidya solution with the mass function that is linear in Eddington-Finkelstein coordinates. Fortunately, this solution-generating technique turns out to preserve the conformal Killing vector field in the original Vaidya metric, leading to separability of null geodesics, and ultimately allows us to develop an analytic formula for the boundary curve of its shadow.  The solution space is parametrized by $\{a, \mu, M_s \}$ where $\{ a, M_s \}$ are the spin and mass parameters of Kerr spacetime in the vanishing $\mu$ limit. Like the Kerr solution, there are regions in the moduli space which do not pertain to black holes. Motivated by phenomenological interests, we focus on the regime of parameters where our solution has event horizons like those of Kerr, with the conformal Killing horizon at a large distance away from the shadow observer and the outer horizon. Thus, our solution serves as a simple model of an accreting Kerr-like geometry not globally but for a finite spatial domain defined by the interior of the conformal Killing horizon. We work in a chart which reduces to the Kerr spacetime in Boyer-Lindquist coordinates in the limit $\mu =0$, and the Vaidya spacetime in Eddington-Finkelstein-like coordinates in the limit $a=0$. Correspondingly, we verified that our shadow formulas reduce consistently to those of Kerr \cite{Grenz,Grenz2} and Vaidya \cite{Solanki} under these limits.

As our main goal here is to seek a metric that is Kerr-Vaidya-like in nature and which also enables us to have an analytic control over its shadow geometry, our starting point and metric construction are agnostic to the form of the energy-momentum tensor. In ordinary Einstein gravity, up to second order in $\mu$, our Kerr-Vaidya-like solution turns out to be sourced by a Type III energy-momentum tensor in Hawking-Ellis classification \cite{Hawking}, some other examples of which were recently noted in \cite{Visser1,Visser2} to admit the interpretation of `spinning null fluids'. Another limitation of our work which deserves further study is that while we could easily identify a null hypersurface $\mathcal{S}_e$ in our Kerr-Vaidya-like solution to be a putative candidate for the event horizon, we were unable to mathematically prove through rigidity theorems that $\mathcal{S}_e$ is formally the boundary of the past of future null infinity unlike the case of the ordinary Kerr solution. However,
there are seminal results in the past literature concerning conformal Killing horizons \cite{SultanaDyer0,SultanaDyer1} which allow us to at least determine $\mathcal{S}_e$ to satisfy local conditions of a putative event horizon. 
In separate limits of vanishing $a$ or $\mu$, $\mathcal{S}_e$ reduces to the event horizons of Kerr and Vaidya solutions respectively. Collectively, these limitations imply that our Kerr-Vaidya-like solution should be regarded as a 
phenomenological toy model of a rotating black hole(-like) spacetime that is accreting mass. Yet most crucially,
it allows us to derive explicit formulas relating between accretion rate and aspects of the shadow geometry.

As reviewed in for example \cite{Oleg}, analytic derivations in cases that allow them complement numerical studies of shadow geometry in general. For example, for Schwarzschild spacetime, the angular diameter of its shadow is $\sim 3\sqrt{3} M_s/R_o$ for a distant observer located at the radial coordinate $R_o$ \cite{Synge}. This numerical value has turned out to be very useful as a guide in the analysis of shadow size and shape in EHT's recent observations \cite{EHTM871,EHTSgr1}. For our shadow analysis here, 
we find a simple power law describing the dependence of the mean radius and asymmetry factor of the shadow on the accretion rate.  The latter describes the departure of the shadow from circularity and has been constrained in M87${}^*$ studies by EHT team \cite{EHTM871}. When applied to the parameters of M87${}^*$ and Sgr A${}^*$, our analysis of shadow geometry appears to indicate that the effect of $\mu$ is very small, and thus provides support for the assumption of using the pure Kerr metric throughout in GRMHD simulations. Our results, in addition, yield an empirical formula that parametrizes the variation of mean radius and asymmetry factor with accretion rate explicitly, and can thus be used to anticipate when backreaction of accretion on the metric may be significant.

Our paper is organized as follows. In Section~\ref{sec:family}, we present the construction of a class of Kerr-Vaidya-like solutions and elaborate on some basic aspects of its geometry and moduli space, followed by a derivation of some analytical formulas for shadow geometry in Section~\ref{sec:Null}. In Section~\ref{sec:portraits}, we present several visual plots of the shadow and examine how the mean radius and asymmetry factor of the shadows vary with various parameters. We also include a brief discussion on recent EHT observations of M87${}^*$ and Sgr A${}^*$ in relation to our model geometry. Finally, we end with some concluding remarks in Section~\ref{sec:discussion}. Appendix~\ref{app:causal} discusses some aspects of the causal structure of our solution, including more detailed discussion 
on conformal Killing horizons and some global aspects of the spacetime geometry. Appendix~\ref{app:energy} presents
some expressions for components of the energy-momentum tensor. 
In Appendix~\ref{app:reference}, for completeness, we develop an aberration formula for observers in another reference frame which, in the zero spin limit, reduces to another class of observers discussed previously in \cite{Solanki} for Vaidya spacetime.

\section{A family of rotating Vaidya-like black hole solutions}
\label{sec:family}
We begin with the Vaidya metric in the coordinates\footnote{The unusual choice of the symbol $w$ to denote the areal radius for this line element is solely due to shortage of conventions for the many different radial coordinates that we'll use throughout this paper.}
\bea
\label{vaidya}
ds^2 &=& - \left( 1 - \frac{2m(v)}{w} \right) dv^2 + 2 dv dw + w^2 \left( d\theta^2 + \sin^2 \theta d\phi^2 \right),
\eea
with the domains $v \in (0, \infty), \, w \in (0, \infty), \theta \in (0, \pi), \phi \in (0, 2\pi)$. 
We note that $m(v)$ is a mass function that can be used to model a time-dependent black hole of which exterior
is described by \eqref{vaidya}. The solution \eqref{vaidya} solves the field equations in ordinary GR with the energy momentum tensor $T^{\mu \nu } = m(v) K^\mu K^\nu, K^\nu \partial_\nu = \partial_w$ which is typically interpreted as that of a null dust moving in the direction of decreasing $w$, 
with the black hole accreting (radiating) mass if $m'(v)$ is positive (negative).

\subsection{Conformal factors, coordinate charts and the Newman-Janis algorithm}

In this work, we restrict ourselves to the case where $m(v) = \mu v$, where $\mu$ is a positive constant. 
In this case, the geometry admits a conformal Killing vector field.  To make this manifest, we make a coordinate transformation as follows.
\be
\label{coord}
v = r_0 e^{T/r_0},  \qquad w = r e^{T/r_0},
\ee
where $r_0$ is a positive constant with dimension of length. This brings \eqref{vaidya} to 
\be
\label{Vai1}
ds^2 = e^{2T/r_0} \left( 
-\left(1 - \frac{2\mu r_0}{r} - \frac{2r}{r_0}  \right) dT^2 + 2 dT dr + r^2 ( d\theta^2 + \sin^2 \theta d\phi^2 )
\right),
\ee
with $T \in (-\infty, \infty), r \in (0, \infty)$. 
In this chart, we can see that a translation in $T$ generates a Weyl rescaling of the metric with $\partial/\partial_T$ defined as the conformal Killing vector. In this form, the metric is conformal to a manifestly static spacetime which can be taken to generate a rotating solution via Newman-Janis algorithm. But first,
we seek a temporal coordinate such that constant time slices are 3-dimensional spatial manifolds. 
Defining 
\be
T = t + \Upsilon (r), \qquad \Upsilon (r) = \int^r d\tilde{R}\,\,  \left(  1 - \frac{2\mu r_o}{\tilde{R}} - \frac{2\tilde{R}}{r_0}  \right)^{-1},
\ee
the line element then reads
\be
\label{VaidyaSch}
ds^2 = e^{\frac{2(t+\Upsilon (r))}{r_0}} \left(  
-\left( 1 - \frac{2\mathcal{M}(r)}{r} \right) dt^2 + \left( 1 - \frac{2\mathcal{M}(r)}{r} \right)^{-1} dr^2 + r^2 ( d\theta^2 + \sin^2 \theta d\phi^2 )
\right) 
 \equiv  \Omega^2 (t,r) ds^2_{static},
\ee
where $t \in (-\infty, \infty )$ and 
$$
\mathcal{M}(r) \equiv \mu r_0 + \frac{r^2}{r_0}.
$$
If we restrict ourselves to the spacetime patch where the conformal Killing vector field 
$\frac{\partial}{\partial t}$ is timelike, then letting $1- 2\mathcal{M}/r >0$ leads to the domain
$$
r \in (R_h, R_c), \qquad R_{h,c} = \frac{r_0}{4} \left( 1 \pm \sqrt{1 - 16 \mu} \right), \qquad \mu < 1/16,
$$
where $R_h$ is the black hole horizon and $R_c$ denotes the conformal Killing horizon\footnote{A more formal definition can be found in \cite{Nielsen}.} -- the 
null hypersurface at which the norm-squared of the conformal Killing vector field vanishes. 
We note that $r = R_h$ is a conformal Killing horizon too, and its interpretation as an event horizon was
recently discussed in \cite{Nielsen} based on an earlier series of results in \cite{SultanaDyer0,SultanaDyer1,SultanaDyer2}.
In \cite{Solanki}, their interpretations were also briefly discussed in the authors' derivation of the shadow formula
for Vaidya spacetime. 

Now, the Schwarzschild limit can be obtained as a double scaling limit as follows.\footnote{ 
The limits that we consider in this paper are smooth limits of metric functions defined specifically
in some coordinate systems. 
See for example \cite{Geroch} for a limiting procedure that covariantizes the process by lifting the
spacetime to a higher-dimensional one with the metric parameter as an auxiliary dimension, and then taking 
its boundary. }
\be
\label{doublescaling}
\mu \rightarrow 0,\,\,\, r_0 \rightarrow \infty,\,\,\, \mu r_o = M_s,
\ee
where $M_s$ is a finite mass parameter equivalent to the ADM mass of the limiting Schwarzschild black hole. 
We now apply the Newman-Janis algorithm\footnote{This algorithm is typically applied 
with the assumption of asymptotic flatness in the generated metric which doesn't hold 
for our solution though. }
to the metric $ds^2_{static}$ which leads to a metric endowed with angular momentum 
\bea
\label{rotating}
ds^2_{static} \rightarrow ds^2_{rotating} &=& - \left( 1 - \frac{2 \mathcal{M}(r) r}{\Sigma} \right) dt^2 - 
\frac{4 \mathcal{M} (r) a r \sin^2 \theta}{\Sigma} d\phi dt \cr
&&\qquad + \left( r^2 + a^2 + \frac{2 \mathcal{M}(r) a^2 r \sin^2 \theta}{\Sigma} \right) \sin^2 \theta d \phi^2
+ \frac{\Sigma}{\Delta} dr^2 + \Sigma d\theta^2,
\eea
where $a$ is the spin parameter and
$$
\Sigma = r^2 + a^2 \cos^2 \theta, \qquad \Delta = r^2 - 2 \mathcal{M}(r) r + a^2.
$$
We will also modify the exponential argument of the conformal factor $\Omega (t,r)$ in \eqref{VaidyaSch} as follows 
\be
\label{Upsilona}
t+\Upsilon (r) \rightarrow t + \Upsilon_a (r),  \,\,\,\, \Upsilon_a (r) \equiv   \int^r  dr\,\,\,\frac{r^2 + a^2}{r^2 - 2\mathcal{M}(r) r + a^2}.
\ee
The full metric then reads 
\be
\label{fullmetric}
ds^2 = e^{\frac{2(t+\Upsilon_a (r) )}{r_0}} ds^2_{rotating},
\ee
with $ ds^2_{rotating}, \Upsilon_a (r)$ being defined in \eqref{rotating} and \eqref{Upsilona} respectively. 
In the scaling limit of \eqref{doublescaling}, 
the line element \eqref{fullmetric} reduces to Kerr spacetime in Boyer-Lindquist coordinates. 

Now, like the Kerr solution where $\mathcal{M}$ is instead just a constant, 
the metric \eqref{fullmetric} has singularities at the roots of $\Delta =0$. To extend the spacetime beyond these singularities, we can perform a coordinate transformation
\be
\tilde{T}= t + \Upsilon_a (r), 
\qquad
\tilde{\phi } = - \phi - a \int^r dr \,\, \frac{1}{r^2 - 2\mathcal{M}(r) r + a^2}
\ee
which leads to
 \bea
\label{tildeT}
ds^2 &=& e^{\frac{2\mu}{M_s} \tilde{T}} \Bigg[ 
-\left( 1 - \frac{2\mathcal{M}(r) r}{r^2 + a^2 \cos^2 \theta} \right) (d\tilde{T} + a \sin^2 \theta d\tilde{\phi })^2 +
2 (d\tilde{T} + a \sin^2 \theta d\tilde{\phi} ) (dr + a \sin^2 \theta d\tilde{\phi }) \cr
\label{KerrV1}
&&\qquad \qquad + (r^2 + a^2 \cos^2 \theta) d\Omega^2
\Bigg],
\eea
In the $a=0$ limit, we recover Vaidya spacetime in the conformally static coordinates of \eqref{Vai1},
whereas the $\mu = 0$ limit (as in \eqref{doublescaling}) takes the metric to that of Kerr in ingoing
Eddington-Finkelstein coordinates. We note that in \eqref{tildeT}, replacing 
$
\mathcal{M}(r) \rightarrow \mu \tilde{T}
$
and removing the conformal factor $ e^{\frac{2\mu}{M_s} \tilde{T}}$ yields the Kerr-Vaidya solution \cite{Dahal}
which evidently isn't equipped with the conformal Kiling symmetry. This leads to non-separability of null geodesics which
would not allow us to solve for the shadow boundary curve analytically.

Our main interest in this class of time-dependent solutions lies in its property of being locally deformable to the Kerr geometry in Boyer-Lindquist coordinates in the $\mu \rightarrow 0,\mu  r_0 \rightarrow M_s$ limit, and, in the limit of $a=0$, to the Vaidya solution in a chart where it's conformally static. \emph{This gives us a model of local geometry that approximates both spacetimes in a coordinate system suitable for deriving the analytical form of the black hole shadow.} For this specific purpose, we work in the $\{ t, r, \theta, \phi \}$ chart (line element in \eqref{fullmetric}), where the spacetime is conformal to a Kerr-like solution in Boyer-Lindquist coordinates. In Section ~\ref{horizons}, we explore the parameter space $\{a, \mu, M_s \}$ in greater detail.

\subsection{Horizons in the solution parameter space}
\label{horizons}

In the ordinary Kerr solution expressed in Boyer-Lindquist coordinates, one can attempt to determine the event horizon
as the hypersurface $r=R_e$, where $R_e$ is the (larger) root of $g^{rr} = 0$, since this equation is equivalent
to setting the Minkowski norm of the normal vector to zero thereby establishing it as a null hypersurface -- a necessary condition. Further, the Killing vector field
\be
\label{KerrKilling}
\xi_K = \frac{\partial }{\partial t} + \Omega_K \frac{\partial}{\partial \phi}, \qquad 
\Omega_K = \frac{a^2}{a^2+R^2_e},
\ee
has vanishing norm on $r=R_e$ (with $\Omega_K$ being the angular velocity of the black hole), and thus $r=R_e$ is a
Killing horizon. These properties satisfy the local conditions for $r=R_e$ to be the event horizon of Kerr, while we can further use rigidity theorems to show that $r=R_e$ satisfies the global notion of the event horizon being the boundary of the causal past of future null infinity. In our case, since we are dealing with dynamical black holes, the event horizon cannot be the Killing horizon defined as above. However, in \cite{SultanaDyer0,SultanaDyer1,SultanaDyer2}, for certain
classes of dynamical black holes which admit conformal Killing vector fields, the authors showed that 
one can similarly locate event horizons through conformal Killing horizons. These were used most recently 
in \cite{Nielsen} to elucidate black hole thermodynamics in the Vaidya spacetime which is the $a=0$ limit of our Kerr-Vaidya-like solution. In Appendix \ref{app:causal}, we discuss more in detail these notions connecting between
conformal Killing horizons and event horizons. 

In a similar vein to finding null hypersurfaces in the Kerr solution of the form $r=$ constant, 
in \eqref{fullmetric}, setting $g^{rr}=0$ yields the following cubic equation in $r$. 
\be
\label{cubic}
-\frac{2\mu}{M_s}r^3 + r^2 - 2M_s r + a^2 = 0.
\ee
In certain regimes of the parameter space of $\{ \mu, a \}$, one could find putative event horizons that necessarily have to satisfy \eqref{cubic} as null hypersurfaces of the form $r=$ constant. 
Consider the root space of the cubic equation \eqref{cubic} of which discriminant reads (henceforth,
we define $a \rightarrow a/M_s$ to be a dimensionless parameter)
$$
\mathfrak{D}=4 \left(  1 - a^2 - 16 \mu + 18a^2 \mu - 27 a^4 \mu^2 \right).
$$
The sign of $\mathfrak{D}$ determines the number of real roots to $g^{rr}=0$. 
As a quadratic equation in $\mu$, we can derive the curves along which 
$\mathfrak{D}=0$, which read
\be
\mu_{\pm} = \frac{-8+9a^2 \pm \left(  4- 3a^2 \right)^{3/2}}{27a^4}.
\ee 
They enclose the region which pertains to three roots $R_i \leq R_e \leq R_a$ which are smoothly connected
(in the limits of vanishing $a,\mu$) to the inner Cauchy horizon of Kerr spacetime($R_i$), the event horizon of Kerr spacetime ($R_e$) 
and the conformal Killing horizon of Vaidya spacetime ($R_a$) respectively.
  
\begin{figure}[h!]
\centering
\includegraphics[width=110mm]{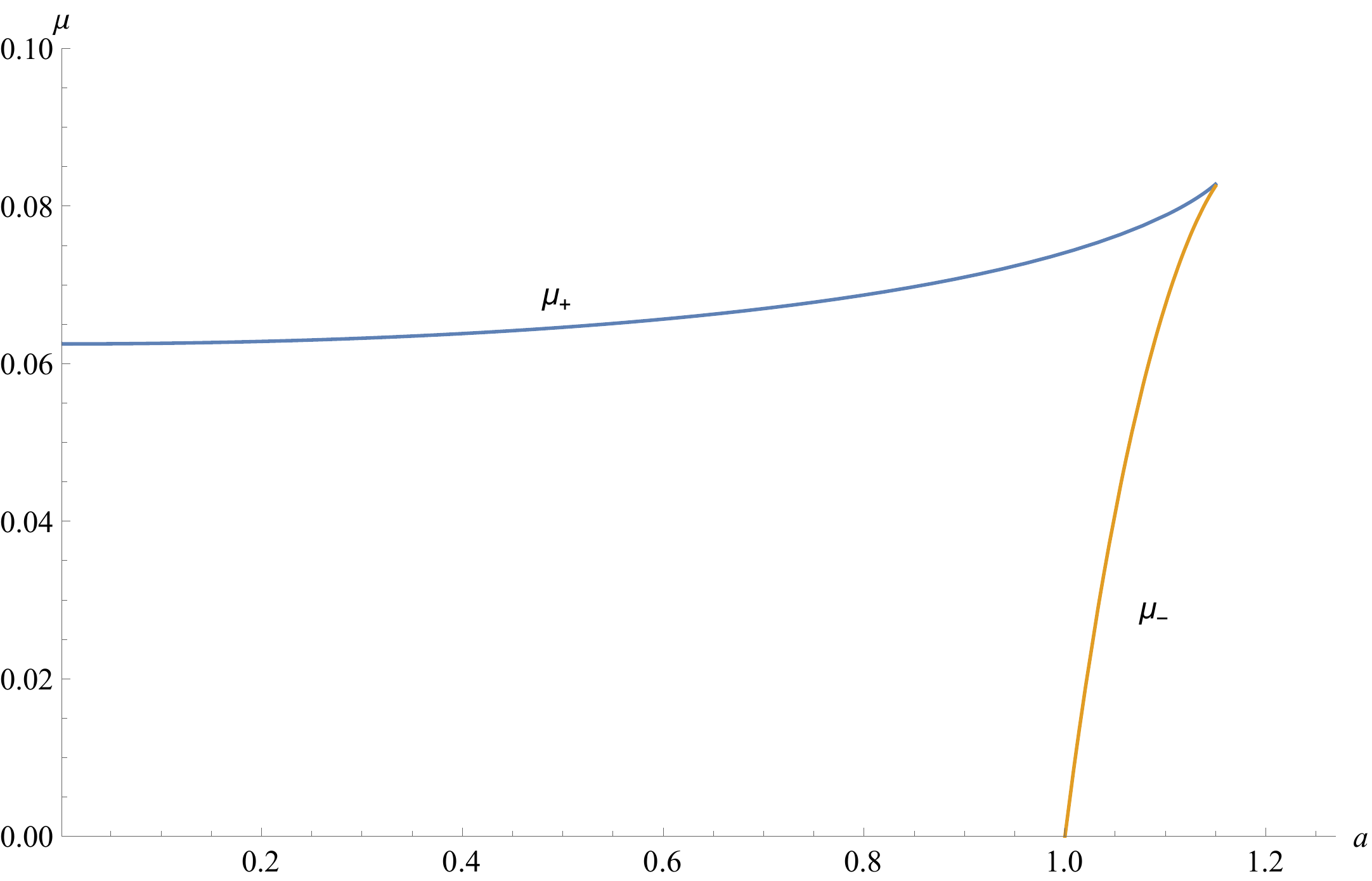}
\caption{Graph depicting the parameter space $(\mu, a)$ of our family of solutions. The curve 
$\mu_+$ begins at $(0,\frac{1}{16})$ and contains solutions where the outer event horizon and conformal Killing horizon are degenerate. The curve $\mu_-$ contains all the extremal solutions with degenerate outer and inner horizons and with a finite conformal Killing horizon radius. Both curves merge at the extremal point
$(\frac{2}{\sqrt{3}}, \frac{1}{12})$ at which there is only one horizon at $r=2M_s$. Our family of solutions can be seen
as parametric deformations of the Vaidya solution (vertical axis) and the Kerr solution (horizontal axis). 
 }
\label{fig:paraplot}
\end{figure}

They intersect at the point (see Fig. \ref{fig:paraplot} )
\be
( a_e, \mu_e) = \left(  \frac{2}{\sqrt{3}}, \frac{1}{12} \right),
\ee
which represents a generalized extremal limit. At any constant $\mu \in (0,\frac{1}{12} )$, 
the upper bound on $a$ is given by taking $\mu = \mu_-(a)$ along which the inner Cauchy and
event horizons coincide. Along $\mu = \mu_+ (a)$, the conformal Killing horizon coincides with 
the event horizon. 

The small $\mu \ll 1$ region of this enclosed segment in Fig. \ref{fig:paraplot} is of closer phenomenological interest to us. Let us consider a generic point in this region. Ordering the roots of \eqref{cubic} as $R_i < R_e < R_a$, we find that up to first few orders in $\mu, a$ :
\bea
R_e &=& M_s \left(  \left(1 + \sqrt{1- a^2} \right)
+ \frac{4+4\sqrt{1-a^2}- 5  a^2 - 3 a^2 \sqrt{1-a^2}    +a^4    }
{2(1-a^2)}u + \mathcal{O}(\mu^2)
\right) \cr
&=& M_s \left[ (2 + 8 \mu ) - (2\mu + 1/2) a + \ldots  \right] \\
R_i &=& M_s \left(  \left(1 - \sqrt{1- a^2} \right)
+ \frac{4 - 4\sqrt{1-a^2}- 5  a^2 + 3 a^2 \sqrt{1-a^2}    +a^4    }
{2(1-a^2)}\mu + \mathcal{O}(\mu^2)
\right) \cr
&=& M_s \left(  \frac{a}{2} + \frac{a^2}{8}  - \frac{\mu a^3}{16} + \frac{a^3}{16} + \ldots  \right) \\
R_{a} &=& \frac{M_s}{2\mu} - 2M_s - 8 \mu M_s + 2a \mu M_s + \mathcal{O}(\mu^2)
\eea
In the limit of vanishing $\mu$,
the radii $R_e, R_i$ are smoothly connected to their corresponding expressions in the ordinary Kerr solution while 
$R_a \rightarrow \infty$. 
\bea
\lim_{\mu \rightarrow 0} R_i &=& M_s \left( 1 - \sqrt{1-a^2} \right) = R^{(\text{Kerr})}_{\text{Cauchy horizon}}, \\
\lim_{\mu \rightarrow 0} R_e &=& M_s \left( 1 + \sqrt{1-a^2} \right) = R^{(\text{Kerr})}_{\text{event horizon}}, \\
\lim_{\mu \rightarrow 0} R_a &\rightarrow& \infty.
\eea
In the $a=0$ limit, $R_i$ goes to zero, $R_e$ reduces to the event horizon of the Vaidya solution, 
whereas $R_a$ reduces to the 
radius of the conformal Killing horizon associated with the conformal Killing vector $\partial_t$ of the Vaidya solution 
\bea
\lim_{a \rightarrow 0} R_i &=& 0, \\
\lim_{a \rightarrow 0} R_e &=& \frac{M_s}{4\mu}\left( 1- \sqrt{1- 16 \mu} \right) = R^{(\text{Vaidya})}_{\text{event horizon}},\\
\lim_{a \rightarrow 0} R_a &=& \frac{M_s}{4\mu}\left( 1+ \sqrt{1- 16 \mu} \right) = R^{(\text{Vaidya})}_{\text{conformal Killing horizon}}.
\eea
In our work, we perform our shadow calculations in the Boyer-Lindquist-like chart of \eqref{fullmetric}
within the domain $r \in [R_e, R_a)$. As shown in Appendix \ref{app:causal}, similar to the case for the ordinary Kerr solution, the conformal Killing vector field 
\be
\label{ckhVector}
\xi = \frac{\partial }{\partial t} + \Omega \frac{\partial}{\partial \phi}, \qquad 
\Omega = \frac{a^2}{a^2+R^2_e},
\ee
has vanishing norm on the null hypersurface $r=R_e$. This is identical in form to Killing vector \eqref{KerrKilling} of Kerr solution which has vanishing norm on Kerr's event horizon. 
As elaborated in Appendix \ref{app:causal}, we can use the results of \cite{SultanaDyer0,SultanaDyer1,SultanaDyer2} 
to interpret $r=R_e$ as the conformal Killing horizon associated with $\xi$ in \eqref{ckhVector}. 
As explained in \cite{SultanaDyer0,SultanaDyer1,SultanaDyer2} , such a null hypersurface satisfies the local conditions of being the putative event horizon of our solutions. The caveat is that we still 
lack some rigidity theorems to rigorously prove that it is an event horizon in the global sense. For a generic point within our domain of interest (region enclosed by curves and axes in Fig. \ref{fig:paraplot} ), our shadow observer is a timelike observer located within $R_e < r < R_a$, i.e. away from the putative event horizon or conformal Killing horizon. 
Since we have been identifying physical interpretations of our Kerr-Vaidya-like solution through limits of metric functions in the chart \eqref{fullmetric}, we would also like our observer to lie in a domain free of any coordinate singularities of \eqref{fullmetric}, in particular, imposing $g_{tt} < 0$ for all polar angles $\theta$ lead to 
\be
R_e < R_h < R_o < R_c < R_a, \qquad R_{h,c} = \frac{M_s}{4\mu} \left( 1 \mp \sqrt{1-16\mu} \right),
\ee
where $R_{h,c}$ are the event horizon and conformal Killing horizon of the Vaidya solution (see \cite{Solanki} for a recent discussion of their physical interpretations).

\section{Null geodesics, photon spheres and shadow formulas}
\label{sec:Null}

\subsection{On null geodesics}
Metrics of the form $ds^2_{rotating}$ in \eqref{fullmetric} admit null geodesics which are separable.
This condition was shown for the Kerr solution in 
for example \cite{Oleg,Grenz}, and for other well-motivated forms of mass function $\mathcal{M}(r)$ in \cite{Naoki}.  Such a property is preserved 
within its conformal class, and in particular by our line element $ds^2$ in \eqref{fullmetric}. To appreciate this, we first recall that for a pair of metrics which are conformally related, say
$
g_{\mu \nu} = \Omega^2(x) \tilde{g}_{\mu \nu},
$
their geodesic equations are related by 
\be
\label{geo1}
\frac{d^2 x^\alpha}{d \eta^2} + \Gamma^\alpha_{\beta \nu} \frac{dx^\beta}{d\eta}\frac{dx^\nu}{d\eta} = 0 
= \frac{d^2 x^\alpha}{d \eta^2} +\tilde{ \Gamma}^\alpha_{\beta \nu} \frac{dx^\beta}{d\eta}\frac{dx^\nu}{d\eta}
-2\Omega^{-1} \frac{d\Omega}{d\eta} \frac{dx^\alpha}{d\eta},
\ee
where $\tilde{\Gamma}^\alpha_{\beta \nu}$ are the Christoffel symbols associated with the metric $\tilde{g}$. 
The last term of \eqref{geo1} can be rewritten as 
\be
\label{geo2}
\frac{d^2 x^\alpha}{d \lambda^2} + \tilde{\Gamma}^\alpha_{\beta \nu} \frac{dx^\beta}{d\lambda}\frac{dx^\nu}{d\lambda} = 0, \qquad \frac{d\lambda}{d\eta} = \Omega^2 (x^\alpha (\eta)).
\ee
Thus, a suitable redefinition of the affine parameter leads to identical forms of the geodesic equations for 
the pair of conformally related metrics. In particular, we expect separability of null geodesic equations just like the Kerr solution or more generally the Kerr-like solutions studied in \cite{Naoki}.

Our solution has a conformal Killing vector field $K\sim \partial_t$ which naturally gives a quantity conserved along its null geodesics. 
In parallel with the notion of energy in the static case, we call this conserved quantity $\tilde{E} = K^\mu P_\mu$, $P_\mu$ being the 4-momentum of a test particle traveling along the geodesic.
Also, the metric components are independent of $\phi$, and we call $L= p_\phi$ the associated conserved quantity. These symmetries motivate the use of
the Hamilton-Jacobi formalism for geodesics where one first defines an auxiliary action $S(\kappa, x^\mu )$ obeying 
\be
\label{HamiltonJacobi}
\frac{\partial S}{\partial \kappa} + \frac{1}{2} g_{\mu \nu} p^\mu p^\nu = 0, \qquad
p^\mu = g^{\mu \nu} p_\nu = g^{\mu \nu} \frac{\partial S}{\partial x^\nu}.
\ee
By virtue of the nature of conserved quantities, we adopt the following ansatz for the action $S$:
\be
S = - \tilde{E}t + L \phi + S_r(r) + S_\theta (\theta) + \frac{1}{2}\mu^2 \kappa,\qquad \mu = -p_\alpha p^\alpha .
\ee
By construction, the 4-momenta $p^\mu = \frac{dx^\mu}{d\eta} = g^{\mu \nu} \partial_\nu S(x)$ satisfies the geodesic equation, with $\kappa = 0$ for null geodesics. 
We find that 
the Hamilton-Jacobi equation \eqref{HamiltonJacobi}
leads to 
\be
-\Delta (r) ( \partial_r S_r )^2 + [ (r^2 + a^2) E - aL ]^2/\Delta (r) = 
( \partial_\theta S_\theta )^2 + ( L - a E \sin^2 \theta )^2/\sin^2 \theta =  \mathcal{K},
\ee
where the constant $\mathcal{K}$ indicates separability. For deriving the shadow formulas, we need the explicit expressions for the 4-momenta which we find to simplify as 
\bea
\label{momenta1}
\frac{dt}{d\eta} &=& \frac{1}{\Sigma (r)} e^{-\frac{2(t+\Upsilon_a (r) )}{r_0}} \left(  
-a(aE\sin^2 \theta - L ) + \frac{(r^2+a^2)P(r)}{\Delta (r)}
\right),
\\
\label{momenta2}
\frac{dr}{d\eta} &=&\pm  \frac{1}{\Sigma (r)} e^{-\frac{2(t+\Upsilon_a (r) )}{r_0}} \sqrt{\mathcal{R}(r)},   \\
\label{momenta3}
\frac{d\theta}{d\eta} &=& \pm \frac{1}{\Sigma (r)} e^{-\frac{2(t+\Upsilon_a (r) )}{r_0}} \sqrt{\Xi (\theta)}, \\
\label{momenta4}
\frac{d\phi}{d\eta} &=& \frac{1}{\Sigma (r)} e^{-\frac{2(t+\Upsilon_a (r) )}{r_0}} 
\left(
-\left(  
aE - \frac{L}{\sin^2 \theta}
\right) + aP(r)/\Delta (r)
\right),
\eea
where 
$$
P(r) \equiv E (r^2 + a^2 ) - a L, \,\, \mathcal{R}(r) \equiv P(r)^2 - \mathcal{K} \Delta (r),\,\,\,
\Xi (\theta) \equiv  \mathcal{Q} + \cos^2\theta \left( a^2 E^2 - L^2/\sin^2 \theta \right),
$$
with $\mathcal{Q}$
being conventionally called the Carter constant defined by 
\be
\mathcal{Q} \equiv \mathcal{K} - (L-aE)^2.
\ee
At this point, we note that apart from the specific form of our mass function $\mathcal{M}(r) = M_s + r^2/r_0$
implicitly contained in $\Delta (r) = r^2 - 2 \mathcal{M}(r) r + a^2$, the 4-momenta expressions 
in \eqref{momenta1} - \eqref{momenta4} are identical to those found in \cite{Naoki} up to the conformal 
factor $e^{-\frac{2(t+\Upsilon_a (r) )}{r_0}}$. This is consistent with the general relations governing conformally related metrics as described in eqns \eqref{geo1} and \eqref{geo2}.

\subsection{Shadow formulas from photon region}
For our analysis of the shadow, we define the constants of motion 
\be
\eta \equiv \frac{\mathcal{Q}}{E^2}, \qquad \xi \equiv \frac{L}{E}.
\ee
Any null geodesics with  $r= R_p$ for some constant $R_p$ leads to the condition
$$
\mathcal{R} (R_p) =  \left( E (R_p^2 + a^2 ) - a L \right)^2
- \mathcal{K} \Delta (R_p)= 0.
$$ 
Further, if the orbit is
unstable then setting $d^2 r /d\eta^2 = 0$ yields
$$ 
\mathcal{R}'(R_p) = 0.
$$
After some algebra, we find that these couple of equations lead to the constants of motion 
\bea
\label{COM1}
a\frac{L}{E}/M^2_s &=& \frac{
4(1+ \mu R^2_p )R^2_p - (3\mu R^2_p + R_p + 1)(R^2_p + a^2 )
}{
-3\mu R^2_p - 1 + R_p
} ,
\\
\label{COM2}
\frac{\mathcal{K}}{E^2}/M^2_s &=& \frac{
( (R^2_p + a^2) - aL/E )^2
}{
R^2_p - 2(1+ \mu R^2_p ) R_p + a^2
}.
\eea
These constants of motion also characterize non-spherical geodesics which asymptotically approach
those confined to spheres defined by $\mathcal{R} (R_p) =\mathcal{R}' (R_p) =0 $. In eqns. \eqref{COM1} and \eqref{COM2}, and henceforth, for notational simplicity, we define $R_p, a$ in units of $M_s$. 
Now consider an observer at the position $(R_{o}, \theta_{inc} )$ and described by an orthonormal
tetrad as follows. 
\bea
\label{tetrad}
&&e_0 = e^{-\frac{\mu T}{M_s}} \frac{(r^2+a^2) \partial_t + a \partial_\phi}{\sqrt{\Sigma \Delta}},
\,\,\,
e_1 = e^{-\frac{\mu T}{M_s}} \sqrt{\frac{1}{\Sigma} } \partial_\theta, \cr
&&e_2 = - e^{-\frac{\mu T}{M_s}}  \frac{\partial_\phi + a \sin^2 \theta \partial_t }{\sqrt{\Sigma} \sin \theta},
\,\, e_3 = -e^{-\frac{\mu T}{M_s}} \sqrt{\frac{\Delta}{\Sigma}} \partial_r.
\eea
As explained in for example \cite{Oleg,Grenz}, this choice leads to $e_0$ being the 4-velocity of 
the shadow observer with $e_0 \pm e_3$ being tangential to the principal null congruences.
(The various expressions in \eqref{tetrad} differ from those in \cite{Oleg,Grenz} by the conformal factor
which we need to take into account for an orthonormalized set of basis vectors.)
The tangent vector of each light ray reaching the observer reads
\be
\label{tangentray}
\frac{d}{d\eta} = p^\mu \partial_\mu  =  \dot{r} \partial_r +  \dot{\theta} \partial_\theta +  \dot{\phi} \partial_\phi 
+  \dot{t} \partial_t = \alpha ( - e_0 + \sin \theta \cos \phi e_1 + \sin \theta \sin \phi e_2 + \cos \theta e_3 ),
\ee
where $\alpha
 = g_{\mu \nu} p^\mu e^\nu_0$.
Equating the coefficients 
after evaluating both sides of \eqref{tangentray} at $(R_{o}, \theta_{inc} )$
yields 
\be
\label{cel}
\sin \Phi = \frac{L_E (R_p)   - a  \sin^2 \theta_{inc} }{\sqrt{\mathcal{K}_E (R_p) } \sin \theta_{inc}    }, \,\,\,
\sin \Theta = \frac{\sqrt{\Delta (R_{o}}) \mathcal{K}_E (R_p)}{R^2_{o} - a L_E (R_p) + a^2}, \,\,\,L_E \equiv \frac{L}{M^2_s E}, \,\,\, \mathcal{K}_E \equiv  \frac{\mathcal{K}}{M^2_s E^2},
\ee
where we have switched to 
a different notation for the celestial coordinates $\phi \rightarrow \Phi, \theta \rightarrow \Theta$ for clarity having derived their eventual expressions.
We note that $\theta_{inc}$ measures the angle between the black hole's 
spin axis as defined by $\theta=0$(the zero locus of $g_{\phi \phi}$) and the observer,
with $e_3$ being parallel to the line of sight connecting the observer to the origin of the Boyer-Lindquist-like 
chart in \eqref{fullmetric}.

The photon region consists of spherical orbits with radii bounded in the domain 
\be
R_p \in ( R_{p,min}, R_{p,max} ), 
\ee
where $R_{p,min}, R_{p,max}$ are defined by setting $\sin \Phi = + 1, -1$ respectively, i.e.
\bea
\label{photonregion}
aL_E(R_{p,min})  &=& a^2 \sin^2 \theta_{inc} + a \sqrt{K_E} (R_{p,min} ) \sin ( \theta_{inc} ), \\
aL_E(R_{p,max})  &=&- a^2 \sin^2 \theta_{inc} - a \sqrt{K_E} (R_{p,max} ) \sin ( \theta_{inc} ) .
\eea
In the vanishing $a$ 
limit, the width of the photon region $(R_{p,min}, R_{p,max} )$ goes to zero, and collapses to 
a single value of $R_p$ which is the photon sphere radius of Vaidya spacetime. In this limit, from \eqref{photonregion}, 
$$
\lim_{a\rightarrow 0} aL_E \rightarrow R^2_p \frac{3+\mu R^2_p - R_p}{R_p-1-3\mu R^2_p} = 0 \Rightarrow
R_p = \frac{1}{2\mu} \left(  1- \sqrt{1-12\mu} \right),
$$
where we have restricted the root to be smaller than the conformal Killing horizon radius. This
is indeed the expression obtained in \cite{Solanki}. Further taking the $\mu =0$ limit yields $R_p = 3$
which is the radius of Schwarzschild photon sphere. In the $a=0$ limit, our expression for 
$\sin \Theta$ in \eqref{cel} reduces to eqn. (41) 
of \cite{Solanki} which is the sine of the angular radius of the Vaidya solution's shadow.

\section{Portraits of the shadow}
\label{sec:portraits}

\begin{figure}[h!]
\flushleft
\begin{subfigure}{.5\textwidth}
  \centering
  \includegraphics[width=\linewidth]{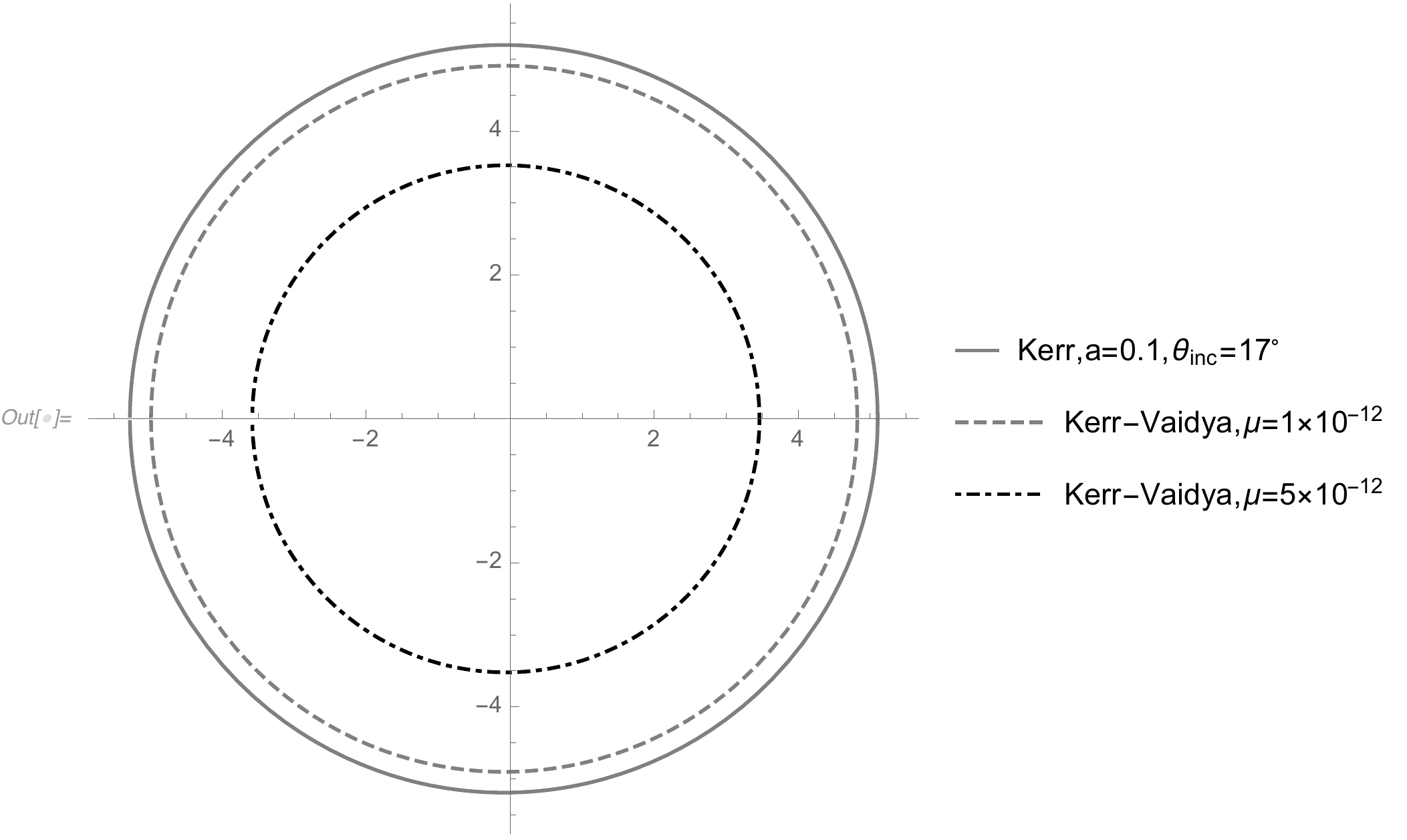}
  \caption{}
  \label{fig:Portrait11}
\end{subfigure}%
\begin{subfigure}{.52\textwidth}
  \centering
  \includegraphics[width=\linewidth]{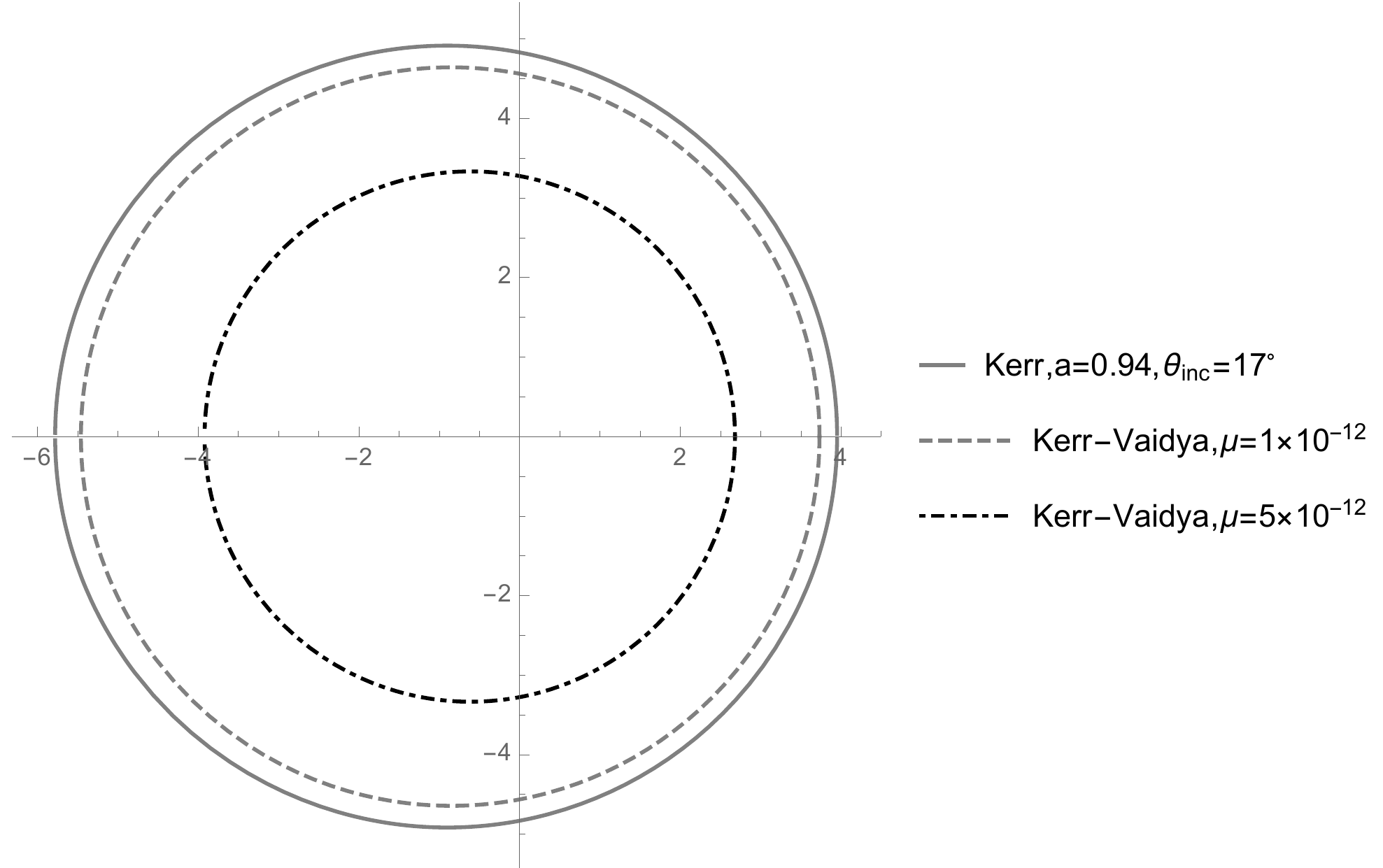}
  \caption{}
  \label{fig:Portrait12}
\end{subfigure}
\begin{subfigure}{.51\textwidth}
  \centering
  \includegraphics[width=\linewidth]{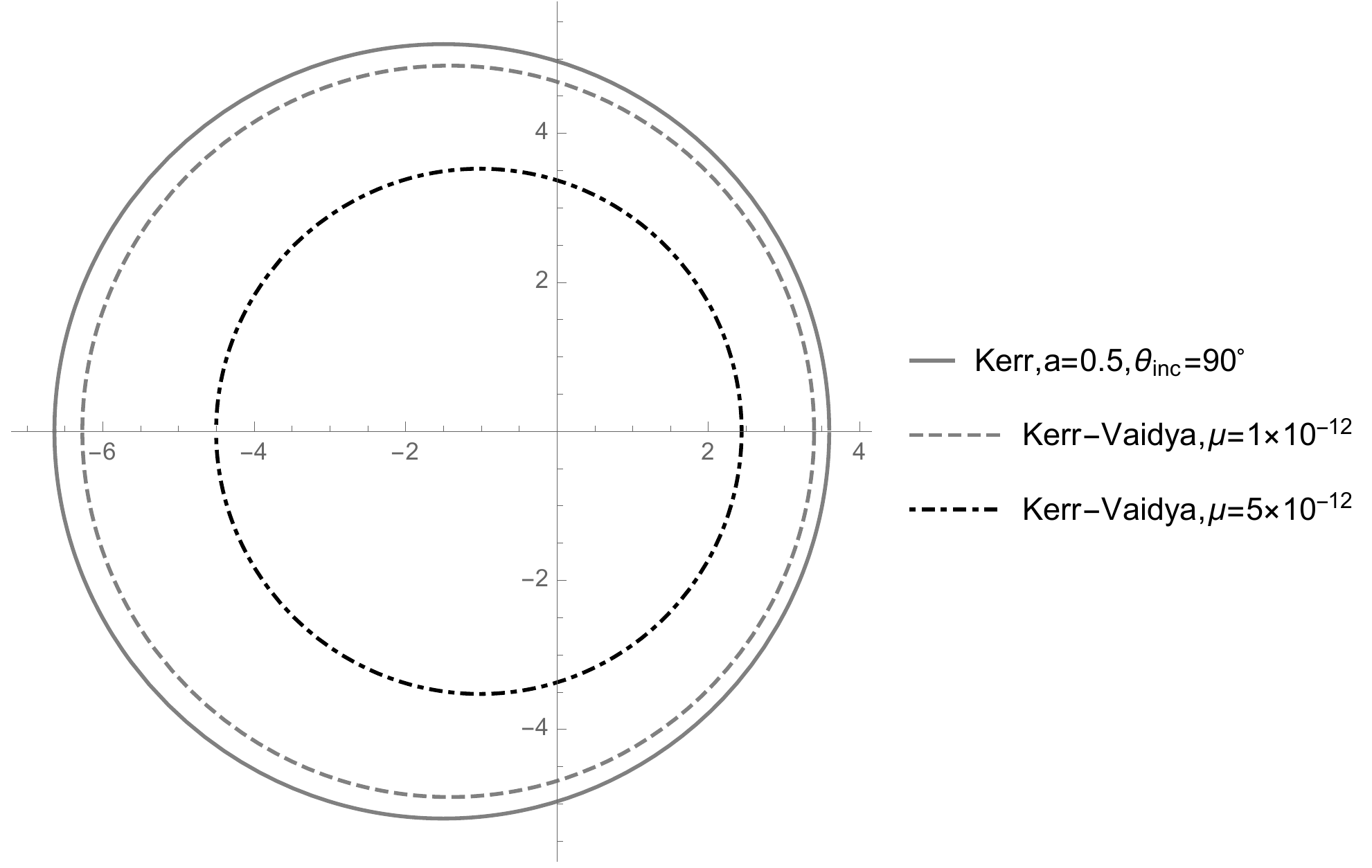}
  \caption{}
  \label{fig:Portrait21}
\end{subfigure}%
\begin{subfigure}{.51\textwidth}
  \centering
  \includegraphics[width=\linewidth]{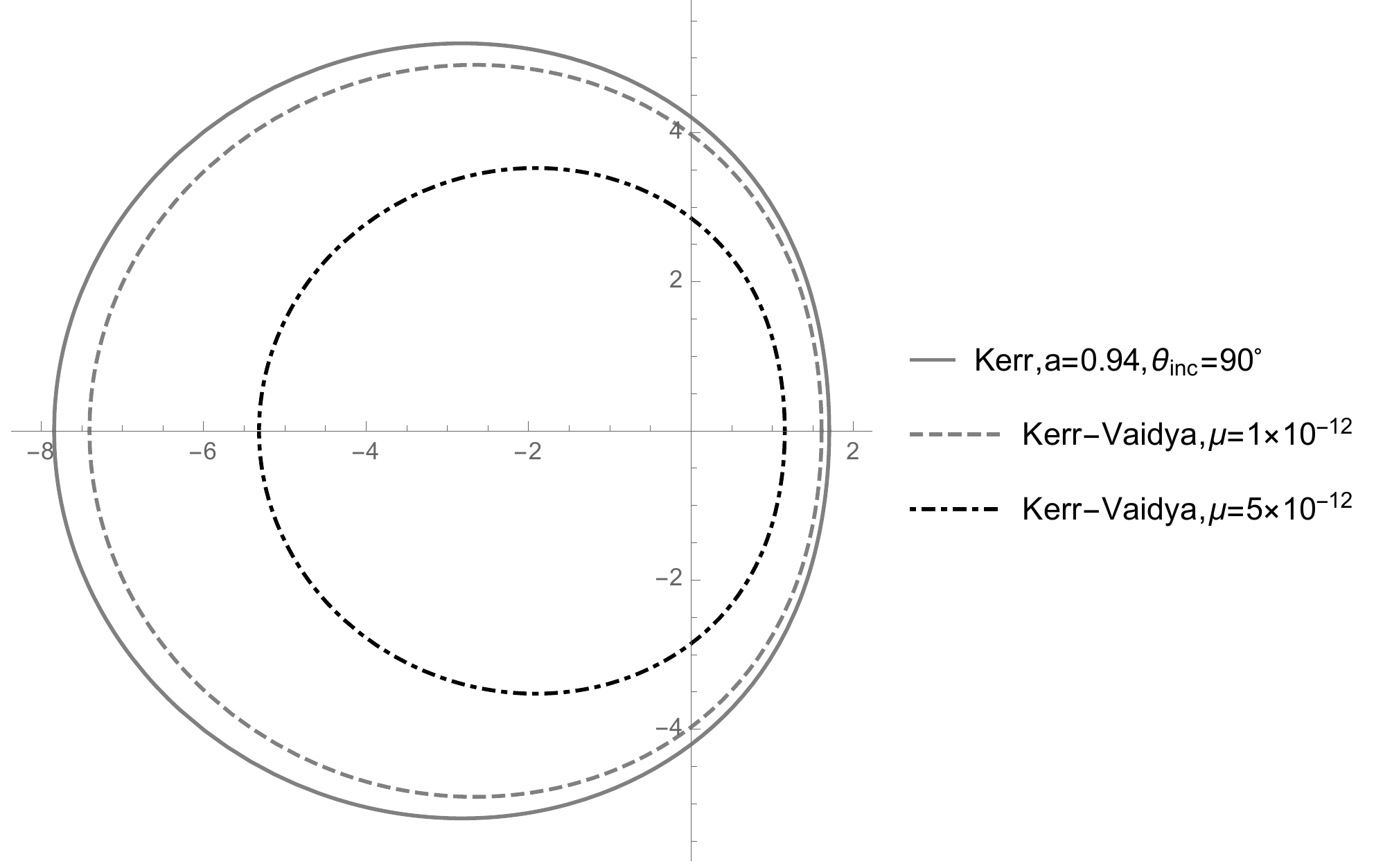}
  \caption{}
  \label{fig:Portrait22}
\end{subfigure}
\caption{Visual representations of the shadow projected upon the observer plane at 
$R_o/M_s = 5.4 \times 10^{10}$
with $\theta_{inc} = 17^\circ$ for (a) and (b), $\theta_{inc} = 90^\circ$ in (c) and (d).  The upper bound on $\mu_b \sim 9.26 \times 10^{-12}$. Horizontal axes
are scaled in units of $M_s$.   } 
\label{fig:PortraitRow}
\end{figure}

In this Section, we discuss geometrical properties of the shadow in more details. We first
recall that the coordinate system describing the shadow observer has a conformal Killing horizon $R_a$ obtained as the largest root of $g^{rr} =0$ in the line element \eqref{fullmetric} which we reproduce explicitly below for convenience.
\bea
\label{Fullmetric2}
ds^2 &=&e^{\frac{2(t+\Upsilon_a (r) )}{r_0}}  \Bigg[  - \left( 1 - \frac{2 \mathcal{M}(r) r}{\Sigma} \right) dt^2 - 
\frac{4 \mathcal{M} (r) a r \sin^2 \theta}{\Sigma} d\phi dt \cr
&&\qquad + \left( r^2 + a^2 + \frac{2 \mathcal{M}(r) a^2 r \sin^2 \theta}{\Sigma} \right) \sin^2 \theta d \phi^2
+ \frac{\Sigma}{\Delta} dr^2 + \Sigma d\theta^2 \Bigg],
\eea
where $\mathcal{M}(r) = M_s + \frac{r^2}{r_0}$ and 
$\Upsilon_a (r) \equiv   \int^r  dr\,\,\,\frac{r^2 + a^2}{r^2 - 2\mathcal{M}(r) r + a^2}$. 
For our shadow observer located at some radial distance $R_o$, we would also like $g_{tt} <0$ for all values of $\theta$, leading to the condition
\be
R_h< R_o < R_c <R_a, \qquad R_{h,c} = \frac{r_0}{4} \left( 1 \mp \sqrt{1-16\mu} \right),
\ee
where $R_h,R_c$ are the event and Killing horizons of the Vaidya solution in the $a=0$ limit. This also implies
that for some fixed $R_o$, we have an upper bound on 
\be
\label{mub}
\mu < \mu_b \equiv M_s \frac{R_o -2M_s}{2R^2_o},
\ee
since we wish to have $R_o < R_c$. For the visual representation of the shadow, we follow the convention used by Johannsen and Psaltis in \cite{Johannsen}. This is essentially the orthonormal tetrad we used
in deriving the shadow formula, with the $x,y$ coordinates being 
\bea
\label{cel2}
x &=& R_o \sin \left(  \Theta (R_p, R_{o} ) \right) \sin \left( \Phi (R_p, \theta_{inc} ) \right), \\
y &=& \pm R_o  \sin \left(  \Theta (R_p, R_{o} ) \right)  \cos ( \Phi (R_p, \theta_{inc} ) ).
\eea
These coordinates parametrize the observer's plane upon which the shadow is projected.\footnote{Another
set of projection coordinates used for plotting the black hole shadow is the $( \alpha,\beta )$ parameters of Bardeen which would not be entirely suitable in our case since our solution is not asymptotically flat. See for example the review of \cite{Oleg} which discusses the relations between Bardeen's impact parameters and others such as stereographic coordinates, etc.}
 From \eqref{cel}, 
noting that $L_E$ and $\mathcal{K}_E$ are odd and even in the spin parameter $a$ respectively, 
one can straightforwardly identify the discrete symmetry 
\be
\label{discrete}
a \rightarrow -a,\,\,\, \theta_{inc} \rightarrow - \theta_{inc},
\ee
which implies in particular that $a<0$ shadows can be obtained from their $a>0$ counterparts
by a reflection in the $y$-axis (for all our shadow plots, we take $a>0$).\footnote{In \cite{Mars}, continuous
symmetries of the shadow modulo Mobius transformations of the celestial sphere were studied for the Kerr-Newman
family of black holes. For the shadow of our black hole solution, we are unable to identify other continuous symmetries
apart from the discrete reflection in \eqref{discrete}. }
In quantifying the shape of the shadow, we follow the work of Johannsen and Psaltis in \cite{Johannsen} who introduced the asymmetry parameter $\mathcal{A}$ to describe departure from circularity. 
\be
\mathcal{A} = 2 \sqrt{\frac{\int^{2\pi}_0 d\alpha\,\,\, ( R - \overline{R} )^2}{\int^{2\pi}_0 d\alpha}},\,\,\,
\tan \alpha = \frac{y}{x}, \,\,\, R \equiv \sqrt{(x-D)^2 + y^2}, \,\,\, D \equiv \frac{|x_{max} + x_{min}| }{2},
\ee
and $\overline{R}= \int^{2\pi} d\alpha\,\, R/ 2\pi$ is the averaged radius projected upon the observer's plane. 
These quantities were mentioned in the EHT paper \cite{EHTM871} for M87${}^*$, and we checked that our shadow geometries for
the background Kerr solution (with $\mu=0$) yield comparable features obtained previously in the work
of Johannsen and Psaltis \cite{Johannsen}. In Figure \ref{fig:PortraitRow}, we picked a few values of $a, \theta_{inc}$ 
for a black hole located at $R_o/M_s \sim 5.4 \times 10^{10}$ (estimated for M87${}^*$) to demonstrate how the shadow curve changes with $\mu$.

\subsection{Scaling laws for variation of $\overline{R}$ and $\mathcal{A}$ with $\mu$}

\begin{figure}[h!]
\flushleft
\begin{subfigure}{.6\textwidth}
  \centering
  \includegraphics[width=0.88\linewidth]{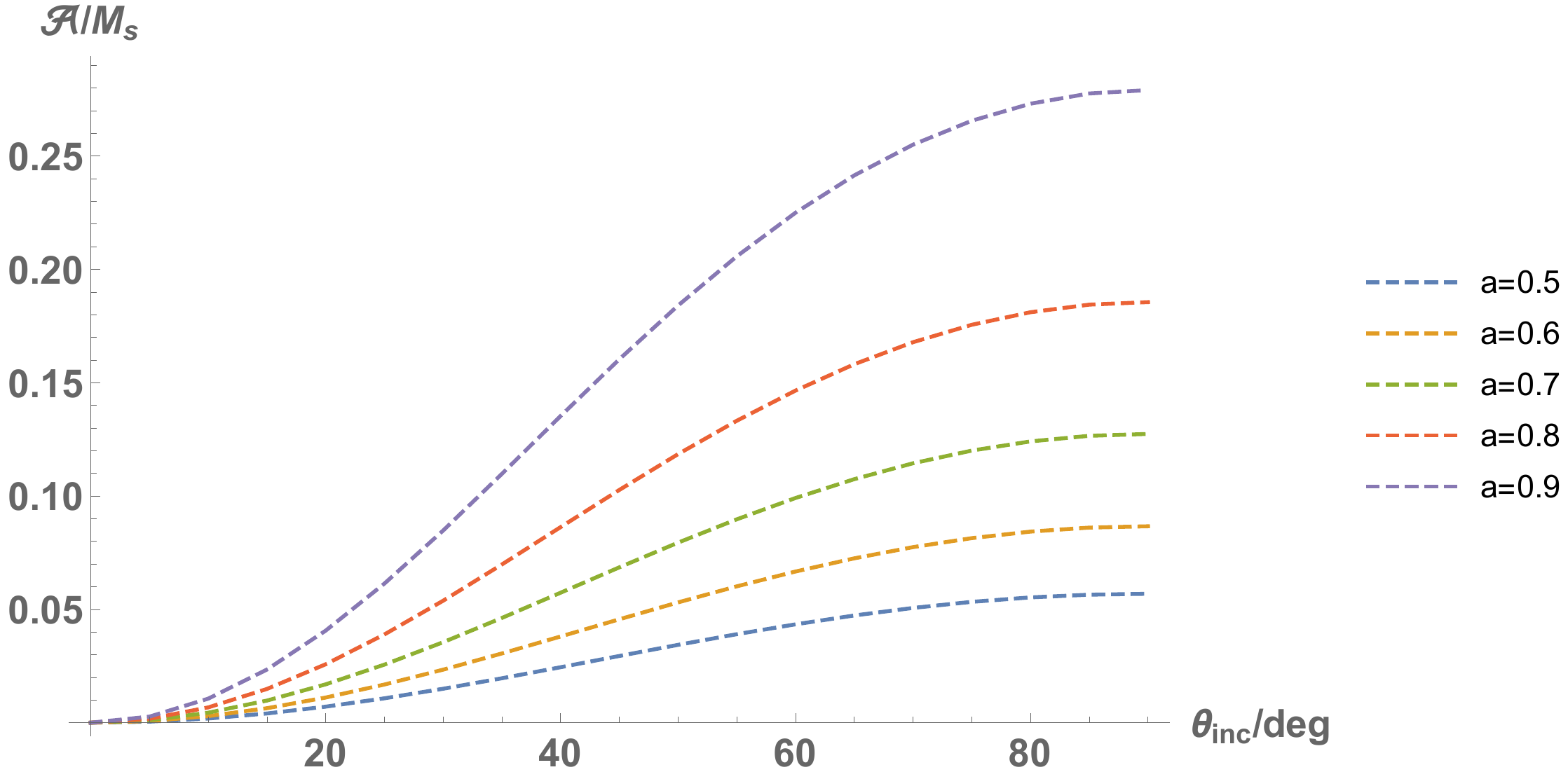}
  \caption{}
  \label{fig:ZeroAsym}
\end{subfigure}%
\begin{subfigure}{.47\textwidth}
  \centering
  \includegraphics[width=\linewidth]{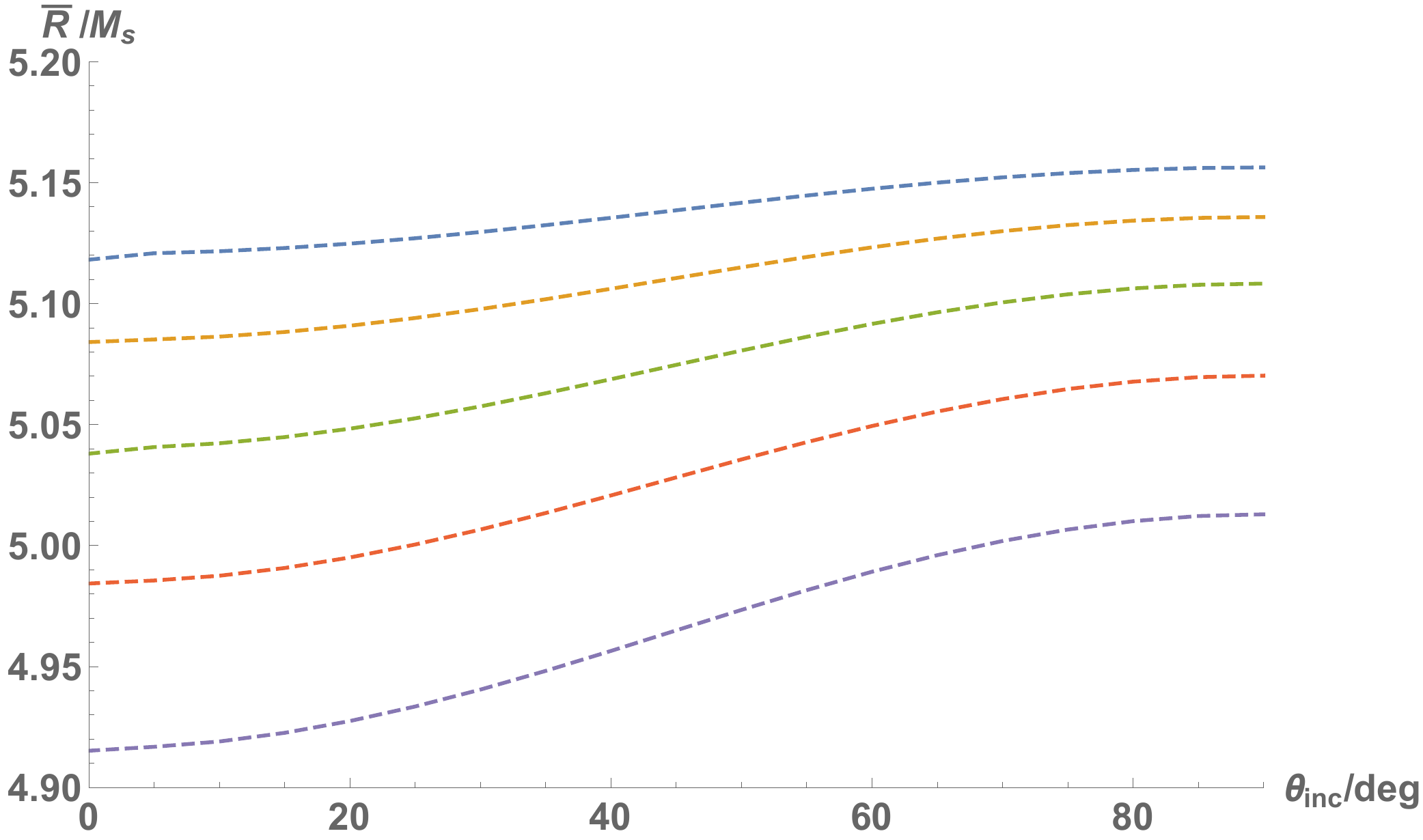}
  \caption{}
  \label{fig:ZeroRad}
\end{subfigure}
\caption{Graphs depicting how $\overline{R}, \mathcal{A}$ vary with angle $\theta_{inc}$, at $\mu = 0$.  We
plot the functions $\overline{R}_o \left( a, \theta_{inc}, R_o \right),  \overline{\mathcal{A}}_o \left( a, \theta_{inc}, R_o \right) $ at five values of $a$ at $R_o/M_s = 5.4 \times 10^{10}$ and $\theta_{inc} = 17^\circ$. }
\label{fig:ZeroQ}
\end{figure}

The parameter space for the shadow geometry is spanned by 
$
\{ a, \theta_{inc}, R_o, \mu \}.
$
Computing the shadow's mean radius and asymmetry factor for a range of parameters, 
we find a simple empirical scaling law that describes the variation of $\overline{R}, \mathcal{A}$ with the accretion rate parameter $\mu$ and other parameters as follows.\footnote{The data-to-model fitting and standard errors of the model parameters was computed using the `NonlinearModelFit' function of Wolfram Mathematica 12.1.1.0 which uses the quasi-Newton method.}
\bea
\label{scalinglaw1}
\overline{R} \approx \overline{R}_o \left( a, \theta_{inc}, R_o \right) \left( 1 - \frac{\mu}{\mu_b (R_o)} \right)^{\kappa_R},\,\,\,
\kappa_R \approx 0.51 \pm 0.01
\\
\label{scalinglaw2}
\overline{\mathcal{A}} \approx \overline{\mathcal{A}}_o \left( a, \theta_{inc}, R_o \right) \left( 1 - \frac{\mu}{\mu_b (R_o)} \right)^{\kappa_A},\,\,\,\kappa_R \approx 0.495 \pm 0.005
\eea
where $\mu_b (R_o)$ is the upper bound \eqref{mub} on the accretion rate allowed by our model for some fixed observer
distance $R_o$, obtained by setting the conformal Killing horizon to be the observer distance.
The dependence on $\mu$ appears as a separate factor independent of the other shadow parameters, with
the functions $\overline{R}_o \left( a, \theta_{inc}, R_o \right),  \overline{\mathcal{A}}_o \left( a, \theta_{inc}, R_o \right) $ describing the radius and asymmetry factor at $\mu =0$. 
The form of \eqref{scalinglaw1} and \eqref{scalinglaw2}  implies that for $\mu \ll 1, R_0 \gg M_s$, 
the fractional decrease in the mean radius and the asymmetry factor that is induced by a non-zero $\mu$ scales approximately as
\be
\label{approximate}
\frac{\delta \overline{R}}{\overline{R}} \approx \frac{\delta \mathcal{A}}{\mathcal{A}} \approx - \mu \frac{R_o}{M_s}.
\ee
In Figure \ref{fig:ZeroQ}, we plot
these functions for a few values of spin at a fixed $R_0$ and $\theta_{inc}$. 
These plots are expectedly similar to 
the corresponding ones presented in \cite{Johannsen} and \cite{Chan}. At higher
spin values, the asymmetry factor and mean radius exhibit a greater range of values over the $\theta_{inc}$ domain. At any $\theta_{inc}$, increasing $a$ increases $\overline{\mathcal{A}}_o$  but decreases $\overline{R}_o$. The form of \eqref{scalinglaw1}, \eqref{scalinglaw2} also implies that the asymmetry factor expressed in units of the mean radius is approximately independent of $\mu$, with 
\be
\frac{\mathcal{A}}{\overline{R}} \approx \frac{\overline{\mathcal{A}}_o }{\overline{R}_o}\left( a, \theta_{inc}, R_o \right).
\ee

\begin{figure}[h!]
\flushleft
\begin{subfigure}{.6\textwidth}
  \centering
  \includegraphics[width=0.9\linewidth]{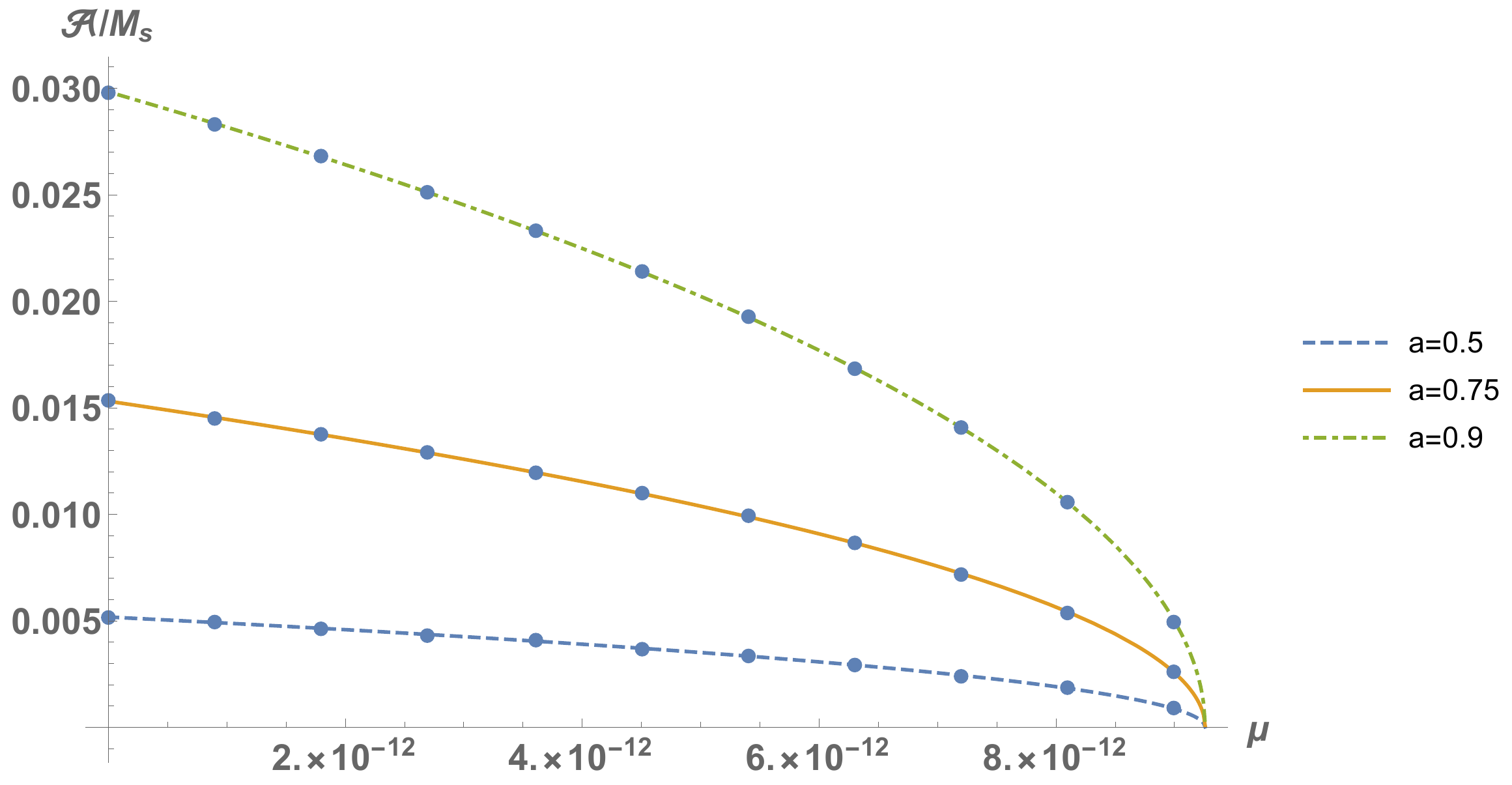}
  \caption{}
  \label{fig:VaryaAsym}
\end{subfigure}%
\begin{subfigure}{.45\textwidth}
  \centering
  \includegraphics[width=\linewidth]{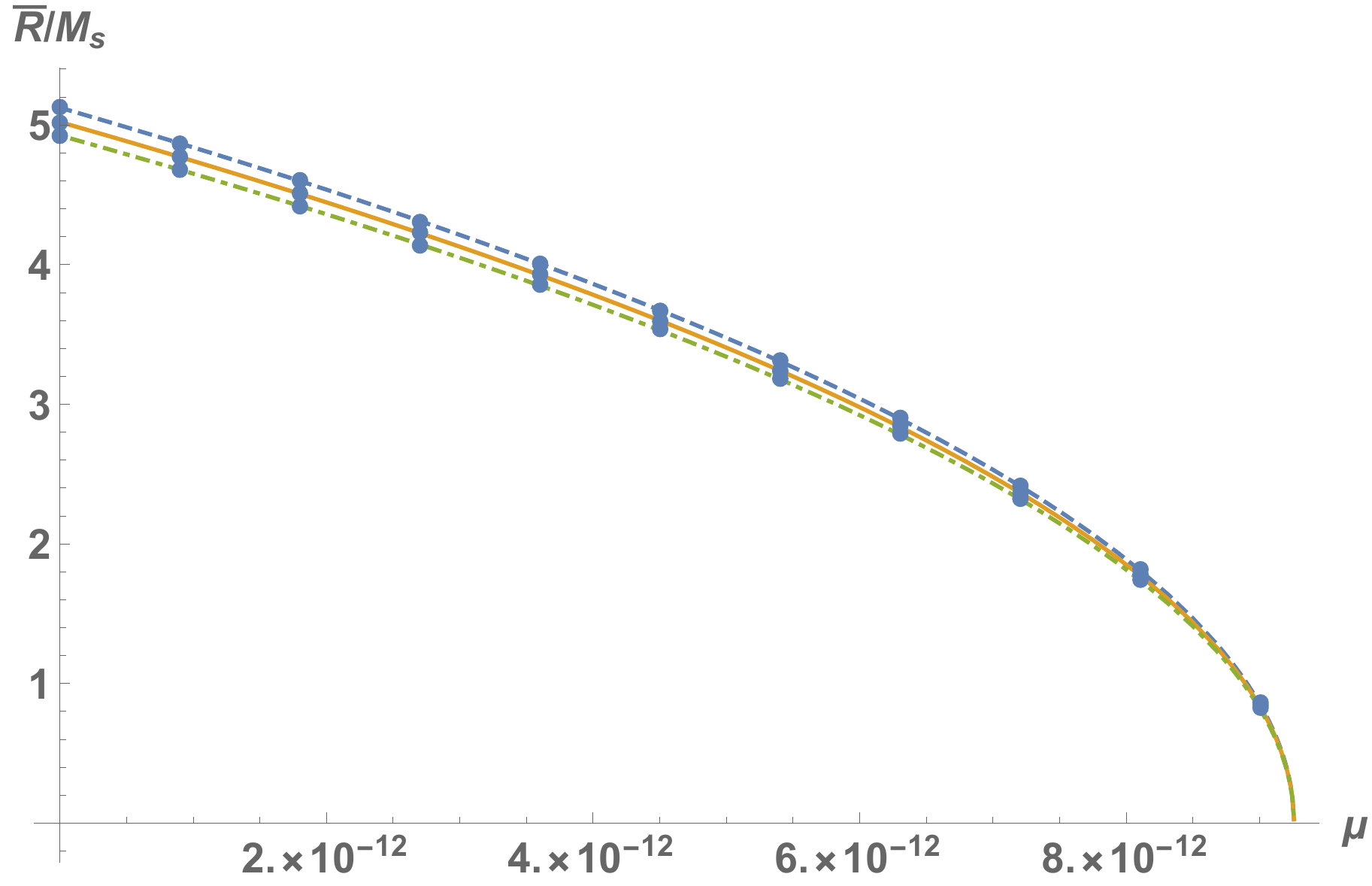}
  \caption{}
  \label{fig:VaryaRad}
\end{subfigure}
\caption{Graphs depicting how $\overline{R}, \mathcal{A}$ vary with $\mu$ for 
$a=0.5, 0.75, 0.9$, with $R_o/M_s = 5.4 \times 10^{10}, \theta_{inc} = 17^\circ$. }
\label{fig:Varya}
\end{figure}

\begin{figure}[h!]
\flushleft
\begin{subfigure}{.61\textwidth}
  \centering
  \includegraphics[width=0.9\linewidth]{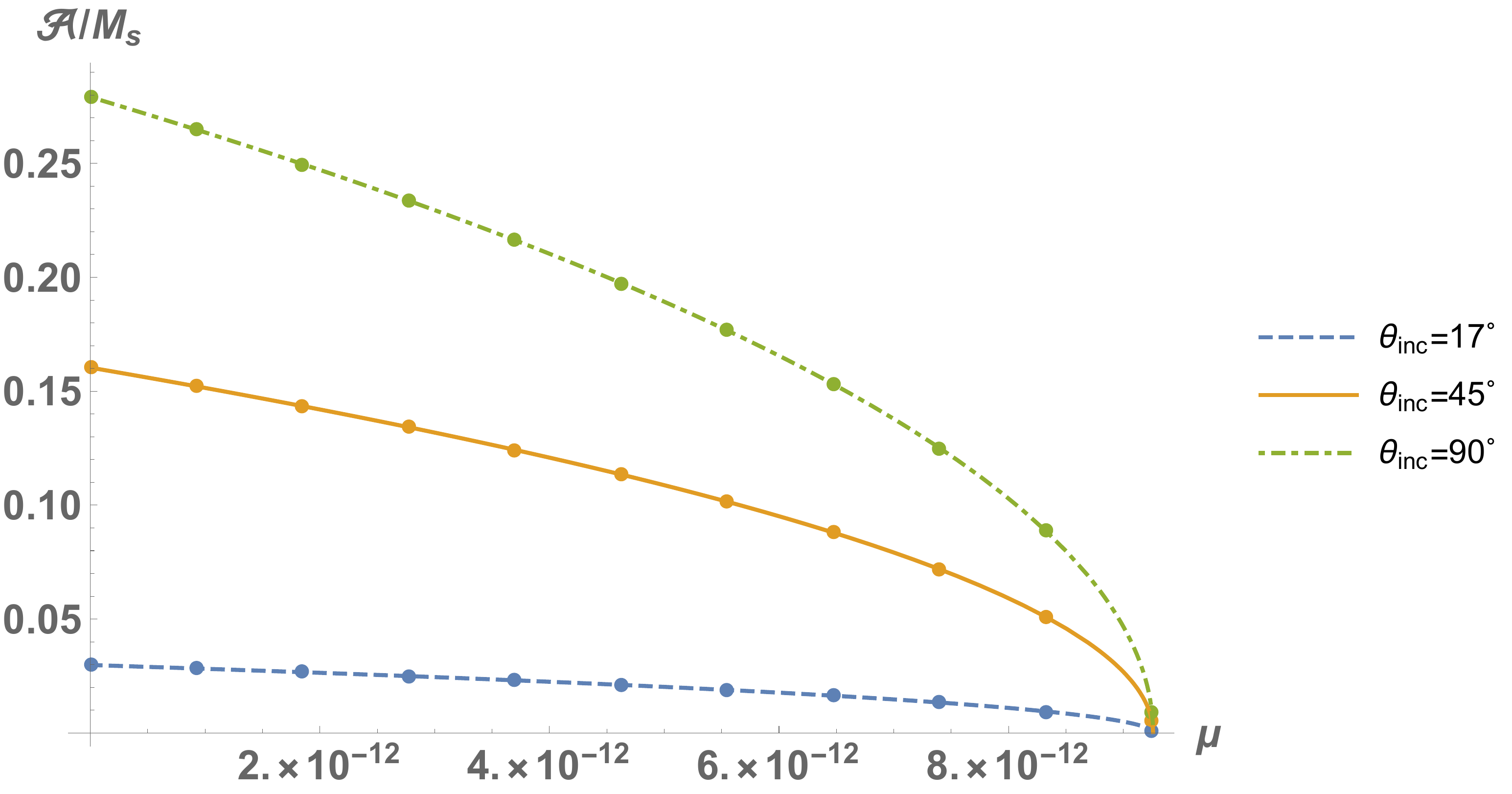}
  \caption{}
  \label{fig:VaryTAsym}
\end{subfigure}%
\begin{subfigure}{.43\textwidth}
  \centering
  \includegraphics[width=\linewidth]{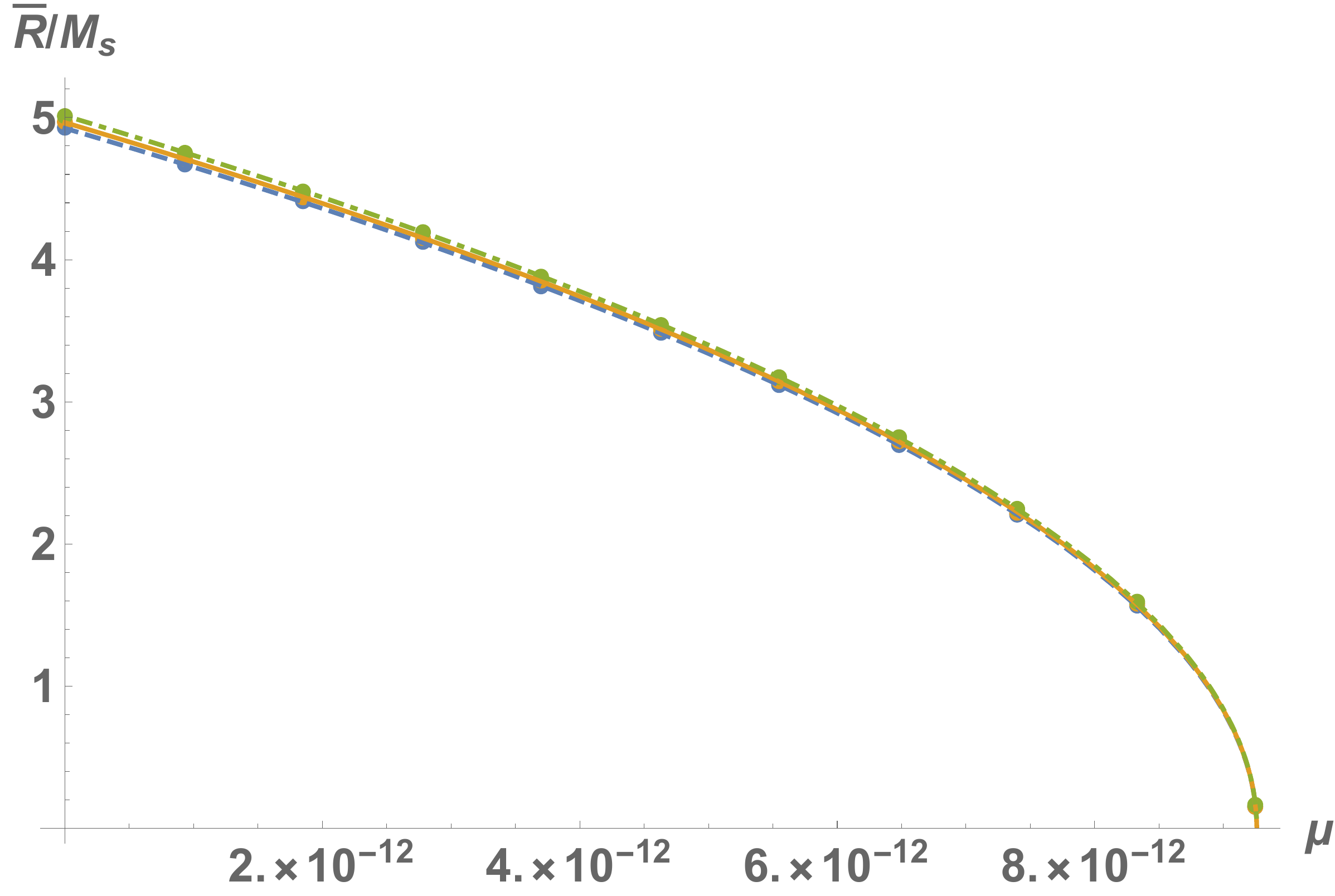}
  \caption{}
  \label{fig:VaryTRad}
\end{subfigure}
\caption{Graphs depicting how $\overline{R}, \mathcal{A}$ vary with $\mu$ for 
$\theta_{inc}=17^\circ, 45^\circ, 90^\circ$, with $R_o/M_s = 5.4 \times 10^{10}, a = 0.9$. }
\label{fig:VaryT}
\end{figure}

\begin{figure}[h!]
\flushleft
\begin{subfigure}{.63\textwidth}
  \centering
  \includegraphics[width=0.88\linewidth]{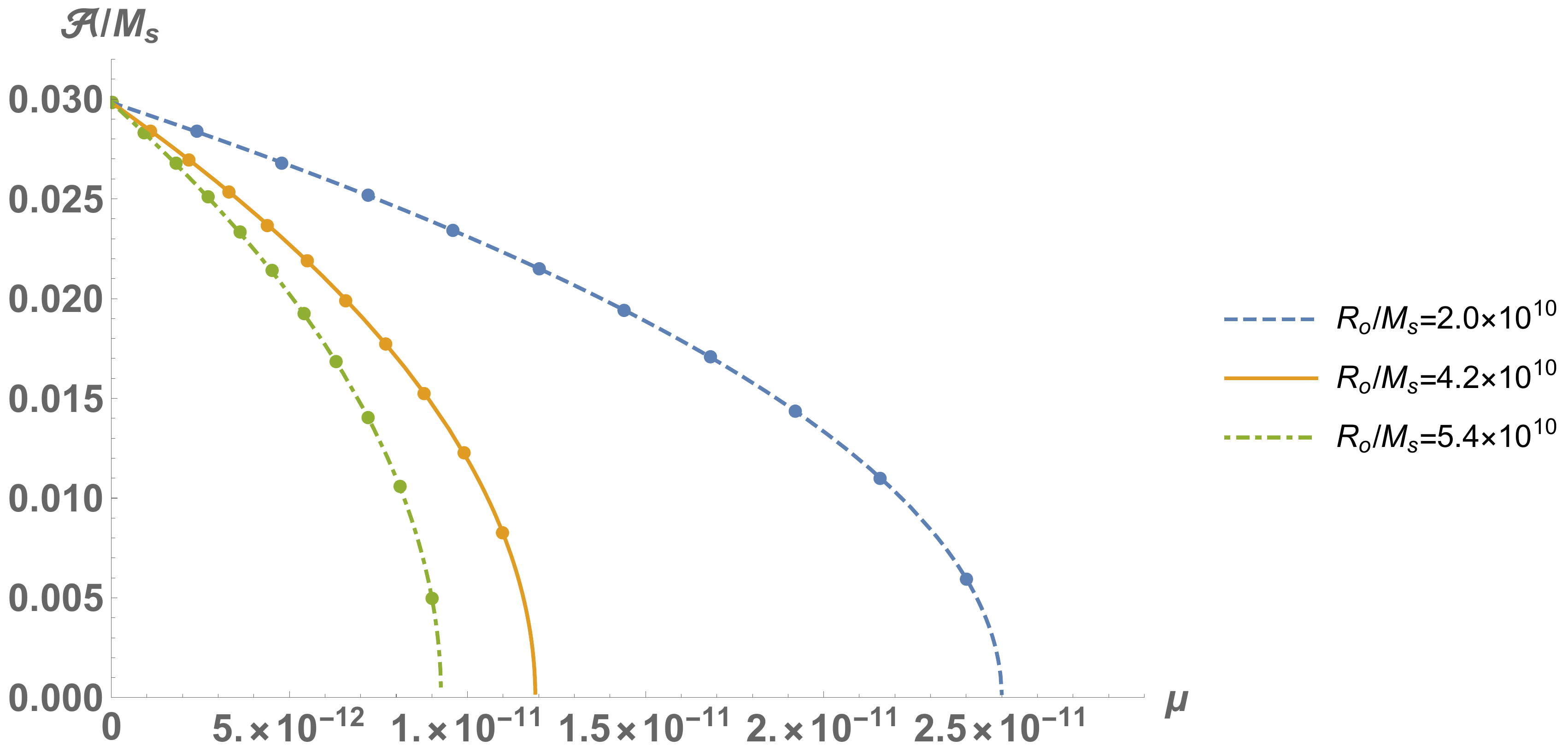}
  \caption{}
  \label{fig:VaryRAsym}
\end{subfigure}%
\begin{subfigure}{.40\textwidth}
  \centering
  \includegraphics[width=\linewidth]{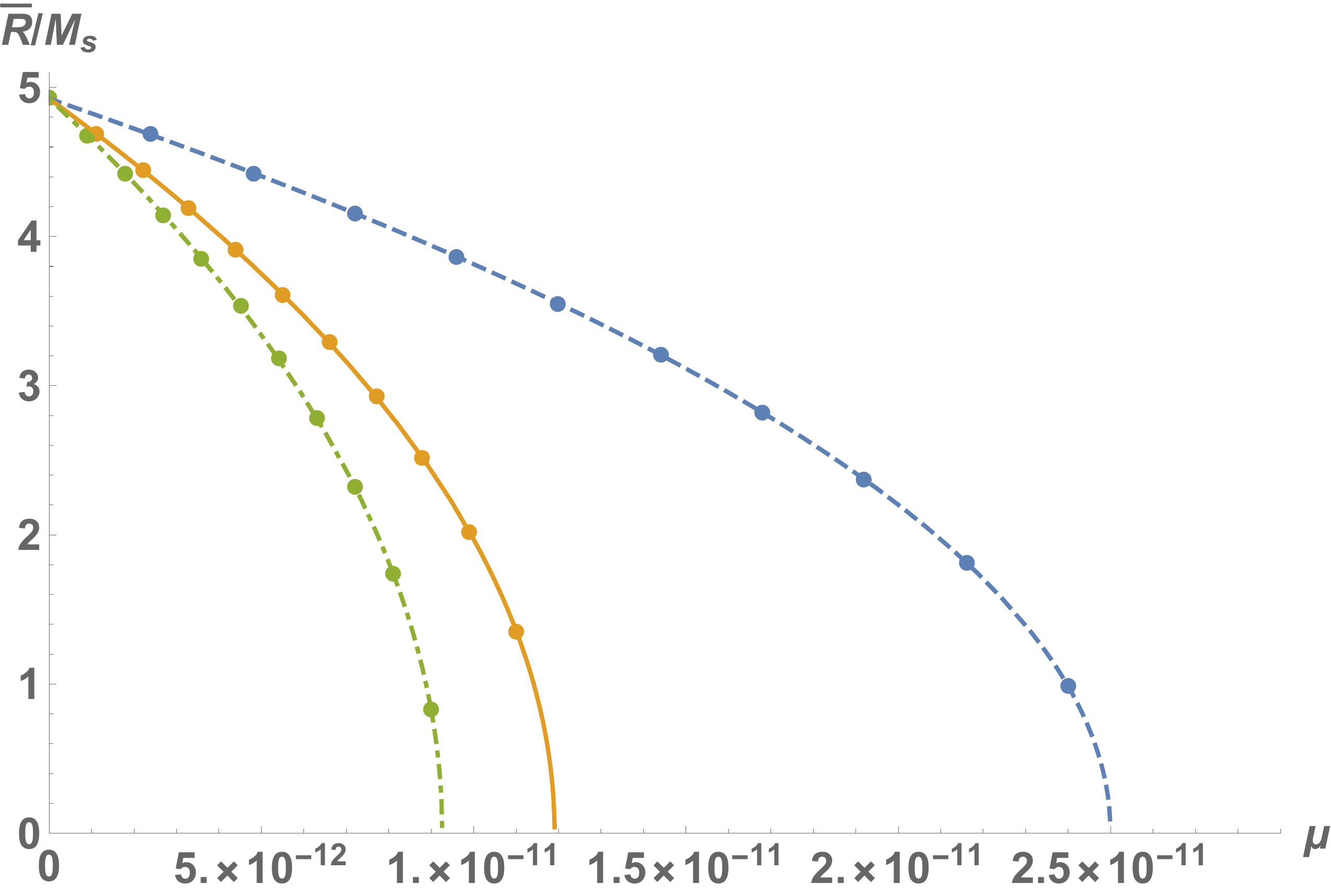}
  \caption{}
  \label{fig:VaryRRad}
\end{subfigure}
\caption{Graphs depicting how $\overline{R}, \mathcal{A}$ vary with $R_o$ for 
$R_0/M_s = 2.0 \times 10^{10}, 4.2 \times 10^{10}, 5.4 \times 10^{10}$ with 
$\theta_{inc}=17^\circ$ and $a = 0.9$. }
\label{fig:VaryR}
\end{figure}

In Figures \ref{fig:Varya}, \ref{fig:VaryT} and \ref{fig:VaryR}, we plot the empirical fitting curves (described by 
\eqref{scalinglaw1} and \eqref{scalinglaw2} ) that depict how $\overline{R}, \mathcal{A}$ vary with the accretion parameter $\mu$
and each of the parameters $\{a, \theta_{inc}, R_o \}$ separately.

%----------------------------------------------------------------

\subsection{On the shadows of M87${}^*$ and Sagittarius A${}^*$ as observed by EHT}

\begin{figure}[h!]
\centering
\includegraphics[width=\linewidth]{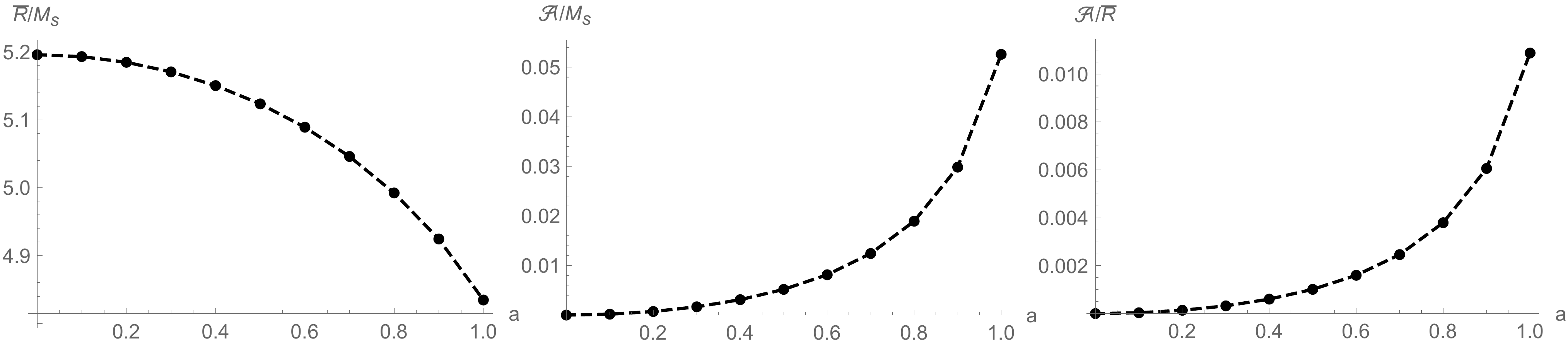}
 \caption{Graphs showing how the mean radius $\overline{R}$, asymmetry factor $\mathcal{A}$ and their ratio vary with $a$, with $R_o/M_s = 5.4\times 10^{10}$ (pertaining to EHT observation of M87${}^*$), $\theta_{inc} = 17^\circ$. }
\label{fig:M87plots}
\end{figure}

The M87${}^*$ black hole was found to be about 16.8 Mpc away, with a mass
 $M_s \simeq 6.5 \times 10^9 M_\odot$\cite{EHTM871}. The ensemble of accretion models used by the EHT team \cite{EHTM871,EHTM875} involved 
mass accretion rates that ranged from about $2 \times 10^{-7} $ to $4 \times 10^{-4}$ times the Eddington rate 
$\dot{M}_{\text{Edd}}$. 
In their work, $\dot{M}_{\text{Edd}} \sim 137 M_\odot / \text{yr}$, and we find that this translates into\footnote{
We note that $\mu$ is the dimensionless mass accretion rate in natural units. To restore its dimension to 
that of mass/time, rescale it by $\mu \rightarrow c^3 \mu / G$, where $G$ is the Newtonian gravitational constant and $c$ the 
speed of light.
} 
$$
\mu_{M87} =  \dot{M} \times \frac{G M_{\odot} }{\text{Yr} \times c^3} \sim \dot{M} \times 1.56 \times 10^{-13} 
\in \left( 4 \times 10^{-18} , 9 \times 10^{-15} \right),
$$
where $\dot{M}$ is mass accretion rate in units of $M_\odot / \text{yr}$. 
This is smaller than the upper bound $\mu_b \sim 9.2 \times 10^{-12}$ for $R_o \sim 16.8$ Mpc. 
Equivalently, for this range of $\mu$, the conformal Killing horizon size falls within
$$
R_c/M_s \sim \left( 5.9 \times 10^{13},  1.16 \times 10^{17} \right),
$$
which lies beyond $R_o/M_s \sim 5.4 \times 10^{10}$. Thus, the observed distance to M87${}^*$ and estimates of mass accretion are well within the domains of validity of our simple model geometry. Now it was estimated in 
\cite{Walker} that the angle of inclination is around $17^\circ$. Corresponding to this value, in Figure \ref{fig:M87plots}, we plot the variation of the shape parameters $\overline{R}, \mathcal{A}$ and their ratio with the spin parameter $a$. The range of values of $\overline{R}$ translates into the shadow angular diameter 
$\sim (36.9 \mu \text{as}, 39.6  \mu \text{as} )$ which is comparable to the measured emission ring diameter in the EHT experiment \cite{EHTM871}. The maximum $\mathcal{A}/\overline{R}$ ratio is about 0.01 which is within limits of the upper bound of $10\%$ indicated in \cite{EHTM871}. The highest $\mu \sim 8.5 \times 10^{-15}$ in the ensemble of models considered in \cite{EHTM871}
translates only to a fractional shift of 0.05 $\%$ in $\overline{R}, \mathcal{A}$. 

Sgr A${}^*$ has been observed to located near the dynamical center of our galaxy at a distance 
$R_o \sim 8$ kpc away, with a dense concentration of mass $M_s \sim 4 \times 10^6 M_{\odot}$.
In contrast to M87${}^*$ where its prominent jet provides robust constraints on source orientation
with respect to the line of sight, fixing it to be $\sim 17^\circ$, there is no such constraint unfortunately on Sgr
A${}^*$ \cite{EHTSgr1}. However, GRMHD models appeared to have favored $\theta_{inc} < 50^\circ$, with accretion rate of order-of-magnitude
$10^{-9} - 10^{-8} M_{\odot}\, \text{yr}^{-1}$. These models were equipped with spin parameter values of $a=0.5, 0.94$  \cite{EHTSgr1}. 
As mentioned in \cite{EHTSgr1}, in the earlier works of Quataert \cite{Quataert} and Baganoff \cite{Baganoff}, the captured accretion rate was estimated to be
$10^{-6} - 10^{-5} M_{\odot} \, \text{yr}^{-1}$ from Chandra observations of thermal bremsstrahlung emission
at the vicinity of the gas capture radius. Most recently, in \cite{EHTSgr5}, a promising model was identified in which $\theta_{inc} \leq 30^\circ$,
and accretion models of $\dot{M}\sim 5.2 - 9.5 \times 10^{-9} M_{\odot}/$yr were examined. 

Even for the accretion rate  $\dot{M} = 10^{-5} M_{\odot} \, \text{yr}^{-1}$ 
this translates into merely
$$
\mu_{sgr} \sim 1.6 \times 10^{-18}.
$$
Like the case of M87${}^*$, this turns out to be smaller than the upper bound $\mu_b \sim 1.2 \times 10^{-11}$ for $R_o \sim 8$ kpc. Equivalently, taking this value of $\mu_{sgr}$, the conformal Killing horizon size is
$$
R_c/M_s \sim 3.2 \times 10^{17},
$$
which lies beyond $R_o/M_s \sim 4.2 \times 10^{10}$. Thus again, both observer distance and (estimated)
mass accretion rates are well within the domains of validity of our simple model geometry. 

\begin{figure}[h!]
\centering
\includegraphics[width=\linewidth]{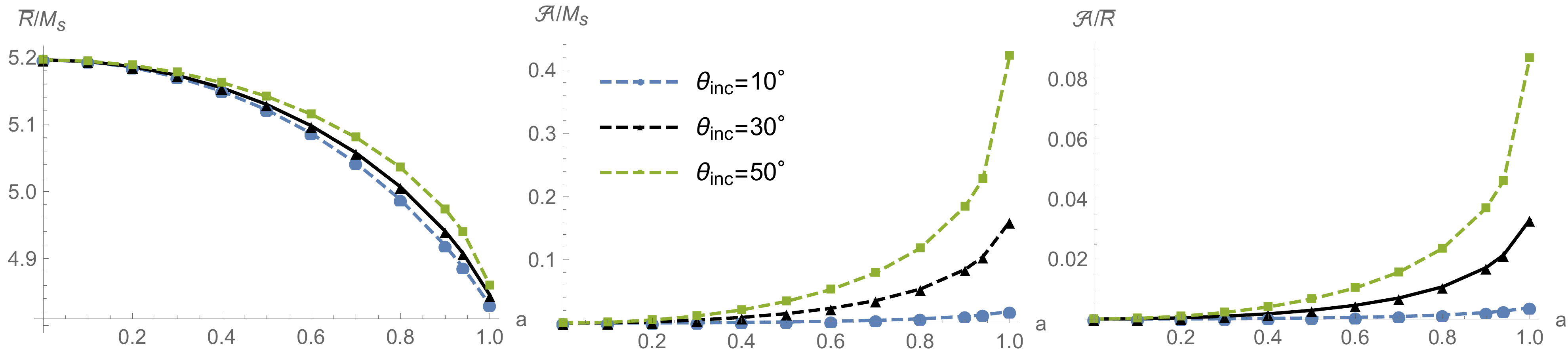}
 \caption{Graphs showing how the mean radius $\overline{R}$, asymmetry factor $\mathcal{A}$ and their ratio vary with $a$, with 
$R_o/M_s = 4.2\times 10^{10}$  (pertaining to EHT observation of Sgr A${}^*$), $\theta_{inc} = 10^\circ,30^\circ,50^\circ$. }
\label{fig:Sagplots}
\end{figure}

In Figure \ref{fig:Sagplots}, we plot the variation of the shape parameters $\overline{R}, \mathcal{A}$ and their ratio with the spin parameter $a$, at a few representative values of $\theta_{inc}$. 
Over the domain of $\theta_{inc} \in (10^\circ, 50^\circ)$,  the range of shadow angular diameters is
$\sim (47.6 \mu \text{as}, 51.2  \mu \text{as} )$ 
which is comparable to the shadow diameter estimate $48.7 \pm 7.0 \mu$as in the EHT experiment \cite{EHTSgr1}. 
The asymmetry factor-to-mean radius ratio $\mathcal{A}/\overline{R}$ ratio increases with $\theta_{inc}$,
and can be as high as $\sim 10 \%$ for $\theta_{inc} =50^\circ$. It would be interesting to study this geometrical signature for 
the EHT's Sgr A${}^*$ shadow image. In \cite{EHTSgr6}, the EHT team mentioned
in passing that the sparse  interferometric coverage of 2017 observations led to significant
uncertainties in circularity measurements which were thus not quantified yet, but future EHT observations with
additional telescopes may place constraints on the circularity. 
Finally, let us mention that the effect of $\mu$ on $\overline{R}, \mathcal{A}$ is even smaller for 
Sgr A${}^*$, inducing only a fractional change of $10^{-6}$ in these quantities.

\section{Discussion}
\label{sec:discussion}

We have presented a study of the shadow geometry for a class of spacetime metric that is Kerr-Vaidya-like in nature.
The family of time-dependent black hole solutions we constructed in this work has well-defined Kerr and Vaidya limits.
Agnostic to the source and the underlying theory, we had conceived of its form starting from
 the Vaidya solution with a mass function
that is linear in Eddington-Finkelstein coordinates since this particular class of solutions furnishes a model for accretion and 
is equipped with a conformal Killing isometry that leads to separability of null geodesics. After expressing it as being
conformal to a solution that is Schwarzschild-like with a radial coordinate-dependent mass function, the Newman-Janis algorithm was applied to obtain a Kerr-like solution that reduces to the Kerr solution in the limit of 
vanishing accretion parameter $\mu$. In real-life applications, the dimensionless accretion rate $\mu$ is expected to be very small, for instance,  for the recent M87${}^*$ and Sgr A${}^*$ observations, the highest model estimates of $\mu$ are of the order $10^{-15}, 10^{-18}$ respectively. Thus, our model geometry can be considered as a small $\mu$-deformation of the Kerr solution that preserves separability of null geodesics, or from a different perpsective, a rotating generalization of the Vaidya solution. For a finite spatial domain, it could act as a simple model of a Kerr-like geometry that takes into account the backreaction of accretion. 
The existence of a conformal Killing vector field allows us to solve for the shadow geometry straightforwardly yet also brings with it the subtlety of a horizon that should be located beyond the shadow observer. Equivalently, at any fixed observer distance, there exists an upper bound to $\mu$ (eqn.\eqref{mub}) for applicability of our model. 

In our study of the variation of the mean radius $\overline{R}$ and asymmetry factor $\mathcal{A}$ with regards to various parameters --- $\{ \mu, a, \theta_{inc}, R_o \}$, we found a simple empirical scaling law of the form in \eqref{scalinglaw1} and \eqref{scalinglaw2}.
In particular, this implies that at small $\mu \ll 1$ and large observer distance $R_o \gg M_s$, the fractional change 
induced by turning on the accretion rate parameter $\mu$ reads simply as 
\be
\label{approx2}
\frac{\delta \overline{R}}{\overline{R}} \approx \frac{\delta \mathcal{A}}{\mathcal{A}} \approx - \mu \frac{R_o}{M_s}.
\ee
To our knowledge, we have not encountered any previous descriptive relations between shadow geometrical features and accretion rate, black hole and observer parameters of the form similar to \eqref{scalinglaw1} or \eqref{approx2}\footnote{In \cite{Kimet}, it was found that increasing the flux of infalling gas into a Schwarzschild black hole (and hence the accretion rate) by increasing an axion-plasmon coupling parameter decreases the size of the shadow which qualitatively agrees with the variation of $\overline{R}$ with $\mu$ of our model. It would be interesting to study if 
the results in \cite{Kimet} could be cast in a similar form as \eqref{scalinglaw1} or \eqref{approx2}.}.GRMHD simulations (e.g. \cite{EHTMag,McKinney,EHTM878}) typically assume the validity of the background metric being purely Kerr in some suitable coordinate system, with the complicated astrophysics of accretion contained within the choice of the energy-momentum tensor. Even for GRMHD simulations involving spacetime metrics motivated by beyond-GR theories (e.g. \cite{EHTSgr6}), the accretion parameter is rarely involved in describing the background spacetime. 
In \cite{Johannsen,Chan} and most recently in \cite{Narayan}, it was found that the accretion details do not appear to influence the shadow geometry which is sensitive only to the background metric. The assumption here is that backreactions of accretion on the metric are insignificant. Indeed, in applying our model to EHT observations 
of M87${}^*$ and Sgr A${}^*$, we found that the most generous estimates of $\mu$ in \cite{EHTM871} and \cite{EHTSgr1} yield fractional changes of
$\overline{R}$ and $\mathcal{A}$ of the order of
$10^{-4}$ and $10^{-6}$ respectively, consistent with such an assumption. In addition, our model yields explicit relations in \eqref{scalinglaw1} and \eqref{scalinglaw2} that describe how, at least in principle, the shadow geometry changes with accretion rate. From 
\eqref{approx2}, it may seem that for situations where the (fractional) experimental uncertainties $\delta \overline{R}_e/ \overline{R}_e, \delta\mathcal{A}_e/\mathcal{A}_e$ are of the order $\mu R_o/M_s$, then metric backreactions due to accretion may be important in analyzing geometrical details of the black hole shadow. 

A limitation of our model geometry is that it is not asymptotically flat. As a model of an effective Kerr-like geometry with accretion backreaction, it is thus valid only for a finite spatial domain. Imposing a stricter condition for $g_{tt}<0$ in \eqref{Fullmetric2}, in our exposition of the black hole shadow analysis, we have taken the observer location $R_o$ to lie within the interior of the sphere defined by the conformal Killing horizon of the limiting Vaidya spacetime, i.e. $R <R_c$ in the coordinate system of \eqref{Fullmetric2}. (For M87${}^*$ and Sgr A${}^*$, $R_c/ R_o \sim 10^3, 10^7$ respectively.) It would be interesting to seek a refinement of our model geometry in one that allows for separability of null geodesics while being asymptotically flat.

\section*{Acknowledgments}
I am grateful to Chong-Sun Chu, Ori Ganor, Petr Horava and Neal Snyderman for sharing with me
their insights on various aspects of gravitational physics over the years. 
I dedicate this work in loving memory of my Aunt 
Mdm Tan Siew Huan to whom I'll always be indebted for her love and guidance.

\appendix
\section{Comments on horizons and causal structure}
\label{app:causal}

Our solution is manifestly a hybrid geometry interpolating between Kerr and Vaidya metrics 
which can be recovered in separate, smooth limits. 
It is characterized by length parameters which reduce to those in Kerr and Vaidya that have been associated 
with horizons. In the main narrative, we have alluded to these parameters as horizon radii as guided by 
the associated 
quantities in the $\mu=0, a=0$ limits. In the following, we furnish more elaborations on these 
geometric notions. 

For a dynamical black hole, its event horizon cannot be described via a Killing horizon, yet
as explained in \cite{SultanaDyer0,SultanaDyer1,SultanaDyer2} and most recently in \cite{Nielsen},
for spacetime geometries which admit a timelike conformal Killing vector field $\xi$, one could, at least
locally, describe a putative event horizon by using the surface generated by the conformal Killing vector
field. In \cite{SultanaDyer0,SultanaDyer1}, the conformal stationary limit surface $\mathcal{S}$ is
defined as the hypersurface on which $\xi$ is null. If such a surface is also a null surface, its generators
are null geodesics as explained in the generic context in \cite{SultanaDyer0,SultanaDyer1} where it was
further shown that since it is a null geodesic hypersurface, it acts as a one-way membrane with the 
future null cone lying on one side of it such that furture directed timelike directions cross the null surface
in the same sense. It was also noted that in spacetimes where $\mathcal{S}$ is not a null surface, 
one can still have the same construction provided there exists a `mixed' conformal Killing field $\tilde{\xi}$
that vanishes on another hypersurface $\tilde{S}$ which is null and interpretable as a conformal Killing horizon. 
In \cite{SultanaDyer0,SultanaDyer1}, it was 
explained how this is analogous to the Kerr solution in which the stationary limit surface (or ergosphere)
is separated from the outer event horizon by an ergoregion. The notion of surface gravity and related
notions of black hole thermodynamics on 
the conformal Killing horizon was also developed most recently in \cite{Nielsen}, where it was shown
how thermodynamical properties of dynamical black holes that admit conformal Killing horizons can be 
elucidated. Notably, the Vaidya spacetime with linear mass accretion which is the $a=0$ limit of our solution was
adopted in \cite{Nielsen} as the concrete illustration of these ideas.

Let us now see how these previous results are applicable to our case. 
Consider our Kerr-Vaidya-like solution in the chart $\{ t, r, \theta, \phi \}$ of \eqref{fullmetric}, where the conformal
Killing vector can be expressed as $\xi = \frac{\partial}{\partial t}$. The conformal stationary limit 
surface $\mathcal{S}$  is the surface $g_{tt}=0$
\be
\mathcal{S} \equiv r^2 + a^2 \cos^2 \theta -2\mathcal{M}(r) r = 0
\ee
One can evaluate the norm-squared of the normal vector $N_\alpha = \partial_\alpha \mathcal{S}$ and straightforwardly verify that this is a timelike surface. In the $\mu = 0$ limit, one can identify $\mathcal{S}$ to be the defining equation 
for the ergosphere of the ordinary Kerr black hole. This motivates us to consider the constant $r$ surfaces in this chart. 
For $r=R_0$ where $R_0$ is some constant,  $N_\mu = \partial_\mu r$, and one can identify the locus of vanishing norm.
\be
N^2 = g^{rr} = e^{-\frac{2\mu}{M_s} \left(t + \Upsilon_a(r) \right)} \left. \frac{r^2 - 2\mathcal{M}(r) r + a^2}{r^2 + a^2 \cos^2 \theta} \right\vert_{r=R_0} = 0,
\ee
which leads to the roots of the cubic equation $r^2 - 2\mathcal{M}(r) r + a^2 = 0$ that we ordered earlier as
$R_i \leq R_e \leq R_a$. The chart is valid for 
the domain $r \in [R_e, R_a)$, since the conformal factor is singular at $R = R_a$ and vanishes at $R = R_e$. 
One can continue past these singularities by using for example the chart in \eqref{tildeT}. 

Now, the surfaces
$R=R_e, R_a$ are null hypersurfaces. Following the prescription of \cite{SultanaDyer0,SultanaDyer1,SultanaDyer2}, 
we seek a conformal Killing vector that is a linear combination of $\partial_t$ and $\partial_\phi$. Writing it as
$$
\tilde{\xi} = \partial_t + \Omega \partial_\phi,
$$
setting $\tilde{\xi}^\alpha \tilde{\xi}_\alpha =0$ yields 
\be
\label{omegaE}
\Omega = \frac{a^2}{a^2 + R_e^2}.
\ee
This expression is identical in form to the angular velocity of rotation
for the Kerr black hole, with $R_e$ being the event horizon radius. \footnote{ In the chart of \eqref{tildeT}, we verified that the same expression for the `mixed' conformal
Killing vector field is obtained. In this case, the $R_e$ appearing in \eqref{omegaE} can be replaced by any root of the 
cubic equation $r^2 - 2\mathcal{M}(r) r + a^2 = 0$ which includes $R=R_i, R= R_a$. }

Since  $\tilde{\xi}^\alpha \tilde{\xi}_\alpha =0$  on the null surface $R=R_e$, the latter is a
geodesic null hypersurfaces as explained in \cite{SultanaDyer0,SultanaDyer1}. 
Locally, it satisfies conditions for the event horizon of a dynamical black hole spacetime. However, to prove that 
$R=R_e$ is the boundary of the region from which null curves cannot escape to infinity, one requires rigidity theorems 
to connect this global notion to the local notion of a conformal Killing horizon as in the ordinary Kerr geometry. In \cite{SultanaDyer0,SultanaDyer1}, generalized versions of strong and weak rigidity theorems were proposed for conformal Killing horizons. However, the conditions specified in them are not satisfied by our model spacetime. 
Specifically, as we discuss in detail in Appendix \ref{app:energy}, the energy-momentum tensor violates the weak energy condition at order up to $\mathcal{O}(\mu^2)$, and this invalidates the proof for a generalized strong rigidity theorem in \cite{SultanaDyer2}. 
Thus, we cannot rely on the results of \cite{SultanaDyer0,SultanaDyer2} for a rigorous proof of $R=R_e$ being the event horizon of our dynamical solution. Nonetheless, as emphasized in \cite{SultanaDyer0,SultanaDyer1,SultanaDyer2}, a conformal Killing horizon is a natural candidate for the event horizon, being a geodesic null hypersurface that acts locally as a one-way membrane for null cones. 

In our work we restrict the observer location $R_o$ in a finite spatial domain of the Boyer-Lindquist-like chart of \eqref{fullmetric} :
$R_h < R_o < R_c$, where $R_h, R_c$ are the roots of $r^2 - 2\mathcal{M}(r) r + a^2 = 0$  in the limit of $a=0$. This ensures that we have $g_{tt} <0$ for all polar angle $\theta$. For this chart, the radial coordinate $r$ lies in the domain
$[R_e, R_a)$, and $R=R_e$ is the only conformal Killing horizon in this interval. In the limit of vanishing $\mu$, it reduces to the outer event horizon of the ordinary Kerr solution. In the limit of vanishing $a$, it reduces to the event horizon of
the Vaidya spacetime as recently identified in \cite{Solanki,Nielsen}. On the other hand,
$R_a \rightarrow \infty$ in the vanishing $\mu$ limit, whereas in the $a=0$ limit, it was identified as a conformal Killing horizon of the Vaidya spacetime in \cite{Solanki,Nielsen} although its physical meaning was not explicitly discussed in these papers.

For the Vaidya spacetime, the conformally static chart in \eqref{Vai1} has a coordinate singularity at the conformal Killing horizon. It can be continued beyond that via \eqref{coord}. 
Similarly, from \eqref{tildeT}, we can perform the coordinate transformation \eqref{coord} 
\be
\label{coord2}
v = r_0 e^{\tilde{T}/r_0},  \qquad w = r e^{\tilde{T}/r_0},
\ee
after which the line element takes the form
\bea
ds^2 &=& - F\left( \frac{w}{v} \right) \left(  dv +  \frac{a v \sin^2 \theta}{r_0}  d\tilde{\phi }  \right)^2 + 
2 \left( dv + \frac{a v \sin^2 \theta}{r_0}  d \tilde{\phi}  \right) \left( dw + \frac{a v \sin^2 \theta}{r_0}  d\tilde{\phi }  \right) \cr
\label{KerrV2}
&& - \frac{2a w \sin^2\theta}{r_0} dv d\tilde{\phi }
+\left(  w^2 + \frac{v^2 a^2 \cos^2 \theta}{r^2_0}  \right) d\Omega^2,
\eea
where 
$$
F\left( \frac{w}{v} \right) = -1 + \frac{2\left(   \mu +  (\frac{w}{v} )^2   \right) \frac{w}{v} }{\left(  \frac{w}{v}  \right)^2 + 
\frac{a^2 \cos^2 \theta}{r^2_0}} - \frac{2w}{v}.
$$
The Vaidya metric in the chart \eqref{vaidya} is obtained in the $a=0$ limit, whereas the 
double scaling limit of \eqref{doublescaling} brings it to Kerr in Eddington-Finkelstein chart. 
The metric in the chart $\{ v, w, \theta, \tilde{\phi} \}$ is convenient for examining the asymptotic infinity of the spacetime. Taking the limit of infinite $w$ yields the following asymptotic form
\be
\label{asymptotic}
\lim_{w \rightarrow \infty} ds^2 \sim -dv^2 + 2 dv dw+ w^2 d\Omega^2 + \frac{2a \mu}{M_s} \sin^2 \theta
\left( w dv d \tilde{\phi }+ v dw d \tilde{\phi} \right).
\ee
At radial infinity there are still non-zero Ricci and Einstein tensor components which read
$
R_{\theta \theta} = -G_{\theta \theta} = - \frac{a^2}{r^2_0} \sin^2 \theta, \,\,\,
R_{\tilde{\phi}  \tilde{\phi}} =  3 G_{\tilde{\phi}  \tilde{\phi}} = -3 \frac{a^2}{r^2_0} \sin^4 \theta,
$
with other components being zero. The asymptotic form of the spacetime shows that at infinity, 
the spacetime geometry is not foliated by two-spheres. For such metrics which lack sufficient symmetry
to be represented by Penrose diagrams of the two-dimensional spacetimes, it was proposed in \cite{Piotr}
that one can still piece together the global causal structure through a set of 2D projections. While we
will not further attempt to invoke such a theory of projection diagrams to analyze our spacetime, in the following,
as a simple probe, we consider the $\theta = \pi/2, \tilde{\phi} = 0$ slice of the metric \eqref{KerrV2} 
and sketch its Penrose diagram
for this 2D spacetime. From this toy Penrose diagram, one can see that in this auxiiary spacetime
with line element 
\be
\label{aux}
ds^2_{aux} = \left( -1 + \frac{2\mu v}{w}  \right) dv^2 + 2 dv dw, 
\ee
the null curve $\frac{w}{v}=\frac{R_h}{r_0}$ represents the boundary of the region containing the entire past of all null rays that could infinity, $R_h = \frac{r_0}{4} \left( 1 - \sqrt{1 - 16\mu} \right)$. Another null curve 
$\frac{w}{v}=\frac{R_c}{r_0}$ with $R_c = \frac{r_0}{4} \left( 1 + \sqrt{1 - 16\mu} \right)$ is the conformal Killing horizon 
of the ordinary Vaidya spacetime described in \cite{Solanki,Nielsen}. We checked that the curvature invariants such as $R_{ab} R^{ab}$ of this auxiliary spacetime are only singular at the $w=0$ locus. We note that in the vanishing $a$ limit, $R_e \rightarrow R_h$, $R_a \rightarrow R_c$. As mentioned earlier in our main narrative, in the chart \eqref{fullmetric}, we restrict the observer's $r$ coordinate
to be in the interval $(R_h, R_c)$ which is precisely the interval in our spacetime solution that leads to $g_{tt}<0$ for all polar angles $\theta$.

\begin{figure}[h!]
\centering
\includegraphics[width=0.4\linewidth]{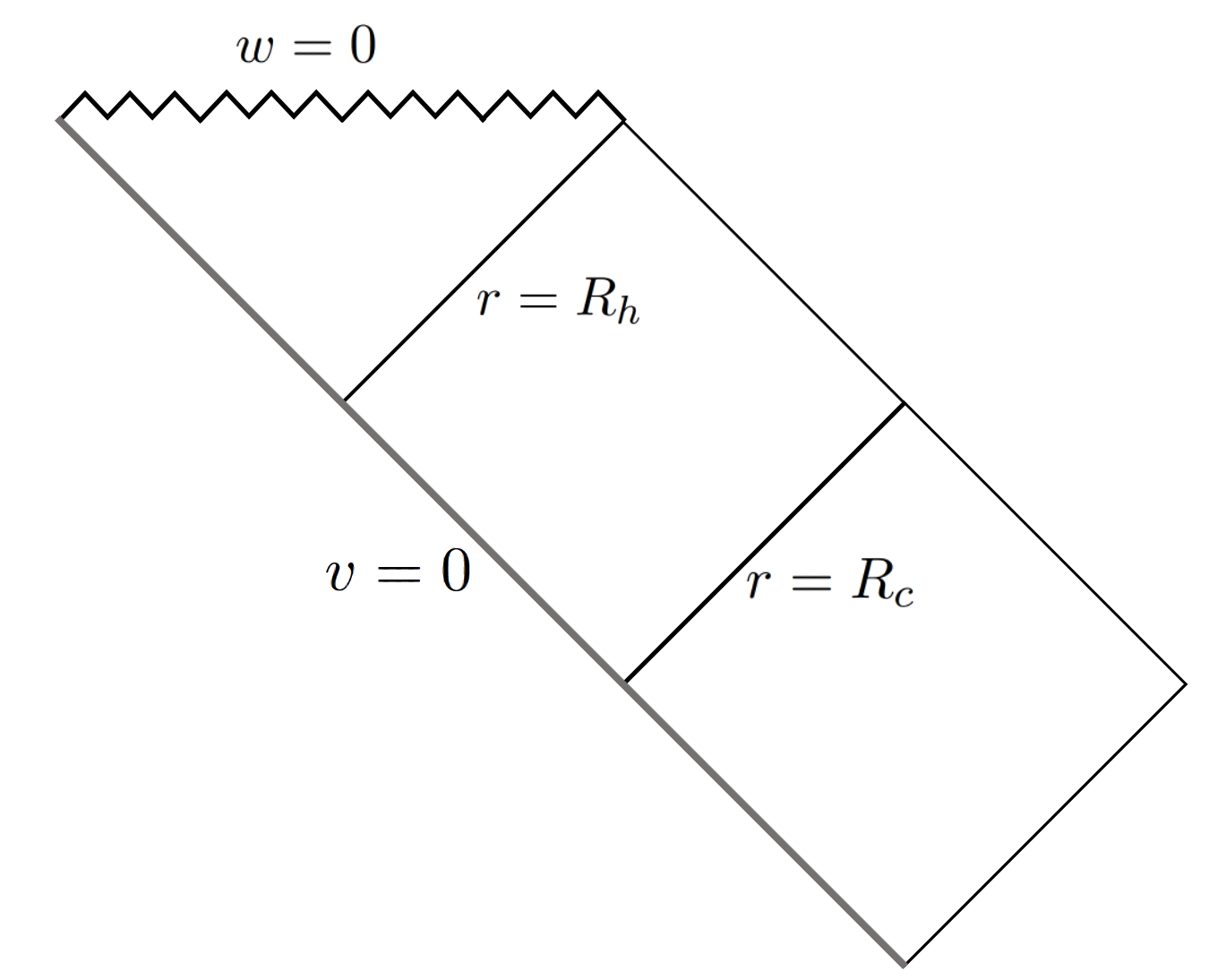}
 \caption{Penrose diagram for the auxiliary 2D spacetime \eqref{aux} which is the $\theta = \frac{\pi}{2}, \tilde{\phi} = 0$ slice of our Kerr-Vaidya-like black hole solution with metric \eqref{KerrV2} . Right boundaries of the two diamonds indicate that this auxiliary 2D spacetime is asymptotically flat but the full 4D spacetime is not. As evident in \eqref{asymptotic}, the asymptotic form of the spacetime is not foliated by two-spheres, and the full causal structure can only be clearly elucidated via different 2D projections, possibly via the techniques introduced in \cite{Piotr}. Nonetheless, this simple auxiliary diagram makes manifest the event horizon interpretation of the curve $r=R_h$ which is the $a=0$ limit of the putative event horizon $r=R_e$.  All curvature invariants of this spacetime admit $w=0$ as the only singular locus. For example, $R^{ab} R_{ab} \sim \frac{8\mu^2 v^2}{w^6}.  $ Our shadow observer is restricted to lie within the central diamond bounded by the horizons.  }
\label{fig:penrose}
\end{figure}

\section{On the energy momentum tensor}
\label{app:energy}

Although we have constructed our spacetime metric in a way that is agnostic to the source, it is instructive to examine the form of the associated energy-momentum tensor in the chart \eqref{KerrV2}. 
In the coordinate system \eqref{KerrV2}, 
to first order in the small dimensionless parameter $\mu$, 
the only non-vanishing component of $T_{\mu \nu}$ is 
\be
8 \pi \mathcal{G} T_{vv} = \frac{2\mu}{w^2}.
\ee
Unfortunately beyond the first order expansion in $\mu$, evaluating the Einstein tensor of our metric yields very complicated expressions for the components of  
 $T_{\mu \nu}$. Up to second-order in $\mu$ , we
find the components to read 
\bea
\label{A1e}
T_{vv} &=& \frac{2\mu}{w} + \frac{A^2 \mu^2}{w^4} \left(
v^2 - 2vw - w^2 + v(v-5w)\cos(2\theta)
\right), \\
T_{vw} &=& \frac{A^2 \mu^2}{2w^4} \left(
(v-w)(2v+w+(2v-w)\cos(2\theta)
\right),  \\
T_{v \theta} &=& \frac{A^2 \mu^2}{2w^2}\left(  -3v -w + (v+w)\cos (2\theta) \right), \\
T_{v \phi } &=& \frac{6\mu^2 Av \sin^2 \theta }{w^2}, \\
T_{ww} &=& \frac{A^2 \mu^2 v^2 \cos^2 \theta }{w^4}, \\
T_{\phi \phi} &=& - \frac{ A^2 \mu^2 v (v-2w) \cos^2 \theta \sin^2 \theta }{w^2}, \\
\label{A2e}
T_{w\phi} &\sim& \mathcal{O}(\mu^3 ), \,\,\, 
T_{\theta \theta} \sim \mathcal{O}(\mu^3 ), \,\,\, T_{\theta \phi} \sim \mathcal{O}(\mu^3 ), 
\eea
where $A \equiv a^2/M^2_s$. These components of the energy-momentum tensor imply that (at least for generic values of $\mu$) the source
cannot be interpreted as a standard null dust of which energy-momentum tensor can typically be cast in the form 
\be
T^{ab} \sim k^a k^b,\qquad k^2 = 0,\, {T^a}_a = 0
\ee
From \eqref{A1e} to \eqref{A2e}, up to and including $\mathcal{O}(\mu^2)$,
we find the trace of the energy-momentum tensor to be
\be
{T^\alpha}_\alpha = - \frac{A^2 \mu^2}{w^4} \left( 
2v^2 - 2vw - w^2 + (2v^2 - 4vw + w^2) \cos(2\theta) 
\right).
\ee
In Hawking-Ellis classification, the nature of the energy-momentum tensor 
in relation to energy conditions can be further elucidated via the eigenvector equation
\be
{T^{\mu}}_\nu n^\nu = \lambda n^\mu. 
\ee
We find that up to and including $\mathcal{O}(\mu^2)$, there are three triply degenerate null eigenvector and one spacelike eigenvector, with the non-zero eigenvalue being the energy-momentum tensor trace ${T^\alpha}_\alpha$,
i.e.
\be
\lambda = \{ 0,0,0, {T^\alpha}_\alpha \}. 
\ee
It can be shown (e.g. \cite{Maeda} )
that a generic Hawking-Ellis type III energy-momentum tensor violates the energy conditions (NEC,WEC,SEC) 
and has no known classical physical origins. More recently in \cite{Visser1,Visser2},  some attempts were made 
at furnishing physical interpretations of type III energy-momentum tensor including examples that the authors argued to 
behave like `spinning null fluids'.

\section{On the reference frame of the shadow observer}
\label{app:reference}

\subsection{In the limit of $a=0$}
In the limit of $a=0$, the tetrad basis \eqref{tetrad} defining the reference frame of our shadow observer reduces to the following.
\be
e_0 = \frac{e^{-\mu \tilde{T}/M_s}}{\sqrt{1- \frac{2\mathcal{M}}{r}}} \partial_t,\,\,\,
e_1 = \frac{e^{-\mu \tilde{T}/M_s}}{r} \partial_\theta , \,\,\,
e_2 = -\frac{e^{-\mu \tilde{T}/M_s}}{r \sin \theta} \partial_\phi , \,\,\,
e_3 = -e^{-\mu \tilde{T}/M_s} \sqrt{\left(  1- \frac{2\mathcal{M}}{r}   \right)} \partial_r.
\ee
In the chart $\{ \tilde{T}, r, \theta, \tilde{\phi} \}$, we replace 
$
\partial_t \rightarrow  \partial_{\tilde{T}},\,\,
\partial_r \rightarrow \frac{\partial \tilde{T}}{\partial r} \frac{\partial}{\partial \tilde{T}}
+ \frac{\partial}{\partial r}
$
which leads to 
\be
\label{vbasis1}
e_0 = \frac{e^{-\mu \tilde{T}/M_s}}{\sqrt{f}} \partial_{\tilde{T}},\,\,\,
e_1 = \frac{e^{-\mu \tilde{T}/M_s}}{r} \partial_\theta , \,\,\,
e_2 = -\frac{e^{-\mu \tilde{T}/M_s}}{r \sin \theta} \partial_\phi , \,\,\,
e_3 = -\frac{e^{-\mu \tilde{T}/M_s}}{\sqrt{f}}  \left( \partial_{\tilde{T}} + f \partial_r \right),
\ee
where $f = 1 -  \frac{2\mathcal{M}(r)}{r} $.
This is the tetrad basis used in \cite{Solanki} for Vaidya spacetime's shadow calculation, relevant
for an observer with 4-velocity $e_0$, at constant $\theta, \phi, r$.

\subsection{Observers in the $\{ v, w, \theta, \phi \}$ chart and aberration formulas}

The coordinate system $\{ v, w, \theta, \phi \}$ avoids the horizons as coordinate singularities. 
In \cite{Solanki}, the shadow observed by an observer with 4-velocity $\sim \frac{\partial}{\partial v}$
was derived using an aberration formula being applied to the shadow angle formula. 

In general, for a pair of reference frames $(S, S')$, 
aberration formulas relating the coordinates of their celestial spheres can be derived from 
the expression \eqref{tangentray} after we express the 4-velocity  $e'_0$ of the $S'$ reference frame as a linear combination of the original tetrad basis components, writing
\be
e'_0 = \frac{e_0 + V^k e_k}{\sqrt{1-v^2}},
\ee
where $\vec{V}$ is the relative 3-velocity of $S'$ observer. 
Consider
an observer of which $e'_0$ is a linear combination of $e_0$ and another basis vector $e_3$. 
Its tetrad basis components read
\be
e'_0 = \frac{1}{\sqrt{1-V^2}} (e_0 + Ve_3),\,\,\, e'_3 =  \frac{1}{\sqrt{1-V^2}} (e_3 + Ve_0), \,\,\, e'_{1,2} = e_{1,2}.
\ee
Taking the inner product between $e'_0, e'_3$ and the tangent vector expression in \eqref{tangentray} in both unprimed and primed coordinates, we obtain the aberration formulas
\be
\label{ab1}
\cos \theta' = \frac{V + \cos \theta}{1 + V \cos \theta}, \qquad \phi' = \phi, \qquad 
V = - \frac{e'_0 \cdot e_3}{e'_0 \cdot e_0}.
\ee
In \cite{Solanki}, the observer with $e'_0 \sim \frac{\partial}{\partial v}$ was also considered. After a coordinate transformation from $\{\tilde{T}, r,\theta, \tilde{\phi} \}$ used for the tetrad basis in \eqref{vbasis1}, one can show that 
\be
e'_0 \sim \frac{\partial}{\partial v} = \frac{1}{\sqrt{1 - \frac{2M_s}{r}}} e^{-\frac{T}{r_0}} 
\left( \partial_{\tilde{T}} - \frac{r}{r_0} \partial_r \right) = \frac{1}{\sqrt{1-V^2}} (e_0 + Ve_3), \,\,
V = \frac{r^2}{r^2 + r r_0 \left(  1 - \frac{2\mathcal{M}(r)}{r} \right)}
\ee
where $e_0, e_1, e_2, e_3$ are as defined in \eqref{vbasis1}. In \cite{Solanki}, the same expression for the relative velocity $V$ was obtained with an aberration relation $ \tan^2 \frac{\theta'}{2} = \frac{1-V}{1+V} \tan^2 \frac{\theta}{2}$ that we verified to be identical to \eqref{ab1}. 

Let us now consider an appropriate $S'$ observer for our Kerr-Vaidya-like geometry. We note that for 
the $S$ observer, its 4-velocity $e_0$ is a linear combination of $\partial_{\tilde{T}}$ and $\partial_{\tilde{\phi}}$, 
or in the Boyer-Lindquist-like chart, a linear combination of $\partial_{t}$ and $\partial_{\phi}$. 
The angular component is such that $e_0 \pm e_3$ are tangential to the principal null congruences of the metric. 
Relating between $v$ and $t$, we choose the following $S'$ observer with 
\bea
e'_0 \sim \frac{\partial}{\partial v} + \frac{r_0 a}{v(r^2 + a^2)} \frac{\partial}{\partial \tilde{\phi}}
&=& \frac{r_0}{v} \left(  1 + \frac{r(r^2 + a^2)}{r_0 (r^2 - 2\mathcal{M}r + a^2)} \right) \partial_t - \frac{r}{v}\partial_r
-\frac{r_0 a}{v(r^2 - 2 \mathcal{M} r + a^2)} \partial_\phi  \cr
&& + \frac{r_0 a}{v(r^2 + a^2)} \frac{\partial}{\partial \tilde{\phi}}.
\eea
This choice of $e'_0$ is also uniquely the one that allows us to write 
\be
e'_0 \sim e_0 + w e_3,
\ee
for some relative 3-velocity $w$. Thus the aberration formulas in \eqref{ab1} apply similarly. 
Recall that in \eqref{tetrad}, the relevant basis tetrad components read 
\be
e_0 = e^{-\mu \tilde{T}/M_s} \frac{(r^2+a^2) \partial_t + a \partial_\phi}{\sqrt{\Sigma \Delta}},
\,\, e_3 = - e^{-\mu \tilde{T}/M_s}  \sqrt{\frac{\Delta}{\Sigma}} \partial_r,
\ee
which allows us to read off the 3-velocity as 
\be
\label{we}
w = \frac{r^2 + a^2}{r^2 + a^2 + \frac{r_0}{r} (r^2 - 2 \mathcal{M} r + a^2 )} .
\ee
In the limit of vanishing $a$, we recover the 3-velocity $v$ for the observer in Vaidya spacetime with $e_0 \sim \frac{\partial}{\partial \tilde{T}}$ as derived in \cite{Solanki}. 
In the limit $\mu \rightarrow 0$, up to leading order, we have
\be
w \approx \mu \left(  \frac{r^2+a^2}{\frac{M_s}{r} (r^2-2M_s r + a^2)}  \right) + \mathcal{O}(\mu^2).
\ee
Thus, this reference frame may be relevant for theoretical situations where the observer's velocity is proportional to the strength of the accretion rate, although arguably not so for realistic EHT observations where the accretion is hardly expected to backreact on the metric significantly to affect the 3-velocity of the shadow observer in such a manner.

\end{document}